\documentclass[12pt]{article}

\usepackage{a4}
\usepackage{pslatex}
\usepackage{hyperref}
\usepackage[latin1]{inputenc}
\usepackage[T1]{fontenc}
\usepackage{graphicx}
\usepackage{amsfonts}
\usepackage{latexsym}

\makeatletter
\def\@sect#1#2#3#4#5#6[#7]#8{\ifnum #2>\c@secnumdepth
  \def\@svsec{}\else 
  \refstepcounter{#1}\edef\@svsec{\csname the#1\endcsname.\hskip0.5em}\fi
  \@tempskipa #5\relax
  \ifdim \@tempskipa>\z@
  \begingroup 
     #6\relax
     \@hangfrom{\hskip #3\relax\@svsec}{\interlinepenalty \@M #8\par}%
  \endgroup
  \csname #1mark\endcsname{#7}\addcontentsline
      {toc}{#1}{\ifnum #2>\c@secnumdepth \else
        \protect\numberline{\csname the#1\endcsname}\fi #7}%
  \else
    \def\@svsechd{#6\hskip #3\@svsec #8\csname #1mark\endcsname
      {#7}\addcontentsline{toc}{#1}{\ifnum #2>\c@secnumdepth \else
        \protect\numberline{\csname the#1\endcsname}\fi #7}}%
  \fi \@xsect{#5}}
\@addtoreset{equation}{section}
\@addtoreset{figure}{section}
\makeatother

\renewcommand\thesection{\Roman{section}}

\renewcommand\theequation{%
  \ifnum \value{section}>0
     \thesection.\arabic{equation}%
  \else
     \arabic{equation}%
  \fi}
\renewcommand\thefigure{%
  \ifnum \value{section}>0
     \thesection.\arabic{figure}%
  \else
     \arabic{figure}%
  \fi}

\def\MSbar{{\overline{\mbox{MS}}}}
\def\mtt{{\rm M}_{t\bar{t}}}
\def\pt{{p_T}}
\def\L{\left(}
\def\R{\right)}

\def\Li2#1{\mbox{Li}_2\left(#1\right)}
\def\GeV{\mbox{ GeV}}
\def\tb{\bar{t}}

\def\ktb{k_{\tb}}

\def\kt{k_{t}}
\def\m#1{{m_{#1}}}
\def\mt{{m_t}}
\def\mb{{m_b}}

\def\mz{{m_Z}}
\def\mw{{m_W}}
\def\mh{{m_H}}

\def\sw{{s_W}}
\def\cw{{c_W}}

\def\gw{{g_W}}
\def\gvf{{g_v^f}}
\def\gaf{{g_a^f}}

\def\gvt{{g_v^t}}
\def\gat{{g_a^t}}

\def\as{{\alpha_s}}
\def\e{\varepsilon}

\def\nn{\nonumber}
\def\Ref#1{Ref.~\cite{#1}}
\def\Refs#1{Refs.~\cite{#1}}
\def\Eq#1{{Eq.~(\ref{#1})}}

\newcommand{\cmq}{{\mt^2\over4\*s_W^2\*\mw^2}}
\newcommand{\hi}{{\mt^2\over4\*s_W^2\*\mw^2}}
\newcommand{\Vtb}{{1\over2\*s_W^2}}

\newcommand{\Amtc}{{{A}_0(\mt^2)}}
\newcommand{\Amzc}{{{A}_0(\mz^2)}}
\newcommand{\Ambc}{{{A}_0(\mb^2)}}
\newcommand{\Amwc}{{{A}_0(\mw^2)}}
\newcommand{\Amhc}{{{A}_0(\mh^2)}}
\newcommand{\Bomtmtmzc}{{{B}_0^Z}}
\newcommand{\Bomtmtmhc}{{{B}_0^H}}
\newcommand{\Bomtmbmwc}{{{B}_0^W}}
\newcommand{\Bosmtmtc}{{{B}_0^t}}
\newcommand{\Bosmbmbc}{{{B}_0^b}}
\newcommand{\Botmtmzc}{{{B}_0^{Z}(z)}}
\newcommand{\Botmbmwc}{{{B}_0^{W}(z)}}
\newcommand{\Botmtmhc}{{{B}_0^{H}(z)}}

\newcommand{\Amt}{{\overline{A}_0(\mt^2)}}
\newcommand{\Amz}{{\overline{A}_0(\mz^2)}}
\newcommand{\Amb}{{\overline{A}_0(\mb^2)}}
\newcommand{\Amw}{{\overline{A}_0(\mw^2)}}
\newcommand{\Amh}{{\overline{A}_0(\mh^2)}}
\newcommand{\Botmtmz}{{\overline{B}_0^{Z}(z)}}
\newcommand{\Botmbmw}{{\overline{B}_0^{W}(z)}}
\newcommand{\Botmtmh}{{\overline{B}_0^{H}(z)}}
\newcommand{\Bosmtmt}{{\overline{B}_0^t}}
\newcommand{\Bosmbmb}{{\overline{B}_0^b}}
\newcommand{\Bomtmtmz}{{\overline{B}_0^Z}}
\newcommand{\Bomtmtmh}{{\overline{B}_0^H}}
\newcommand{\Bomtmbmw}{{\overline{B}_0^W}}
\newcommand{\DBomtmtmz}{{{d\over dp^2}B_0^Z\Big|_{p^2=mt^2}}}
\newcommand{\DBomtmtmh}{{{d\over dp^2}B_0^H\Big|_{p^2=mt^2}}}
\newcommand{\DBomtmbmw}{{{d\over dp^2}B_0^W\Big|_{p^2=mt^2}}}
\newcommand{\Cotmtmtmz}{C_0^{Z}(z)}
\newcommand{\Cotmbmbmw}{C_0^{W}(z)}
\newcommand{\Cotmtmtmh}{C_0^{H}(z)}
\newcommand{\Cosmtmtmt}{C_0^t}
\newcommand{\Cosmbmbmb}{C_0^b}
\newcommand{\Cosmtmtmz}{C_0^Z}
\newcommand{\Cosmbmbmw}{C_0^W}
\newcommand{\Cosmtmtmh}{C_0^H}
\newcommand{\Dotz}{D_0^{Z}(z)}
\newcommand{\Dotw}{D_0^{W}(z)}
\newcommand{\Doth}{D_0^{H}(z)}

\newcommand{\genfacs}{N\*{\mt^2\over s^2}\*{z^2\over1-\beta^2\*z^2}}
\newcommand{\genfacschi}{N\*{\mt^4\over s^2}\*{z^2\over1-\beta^2\*z^2}}
\newcommand{\genfacself}{{2-N^2\*(1-\beta\*z)\over N\*(1-\beta^2\*z^2)}}
\newcommand{\genfacvert}{{2-N^2\*(1-\beta\*z)\over N\*(1+\beta\*z)^2}}
\newcommand{\genfacbox}{{N^2\*(1-\beta\*z)-2\over N\*(1-\beta^2\*z^2)}}

\newcommand{\ra}{\rightarrow}
\newcommand{\chifacs}{\gat^2}
\newcommand{\chifac}{{\gat^2\over\mz^2}}

\begin{document}
\thispagestyle{empty}
\begin{flushright}
  CERN-PH-TH/2006-222\\
  TTP06-28\\
  SFB/CPP-06-49
\end{flushright}
\vspace*{3cm}
\begin{center}
  {\Large\bf
    Electroweak effects in top-quark pair production \\[0.2cm] 
    at Hadron Colliders
    }\\
  \vspace*{1cm}

  J.H.~Kühn$^a$, A. Scharf\,$^a$ 
  and P. Uwer\,$^b$\footnote{Heisenberg Fellow}\\
  \vspace*{0.5cm}
  {\em $^a$Institut für Theoretische Teilchenphysik,
    Universität Karlsruhe\\ 76128 Karlsruhe, Germany}\\
  {\em $^b$CERN, Department of Physics, Theory Unit,\\
    CH-1211 Geneva 23, 
    Switzerland}    
\end{center}
\vspace*{1.5cm}
\centerline{\bf Abstract}
\begin{center}
  \parbox{0.8\textwidth}{
    Top-quark physics plays an important r\^ole at hadron colliders
    such as the Tevatron collider at Fermilab or the upcoming 
    Large Hadron Collider (LHC) at CERN. 
    Given the planned experimental 
    precision, detailed theoretical predictions are 
    mandatory. In this article we present analytic results for the
    complete electroweak
    corrections to gluon induced top-quark pair production, completing
    our earlier results for the quark-induced reaction. As an
    application we discuss top-quark pair production at Tevatron and
    at LHC. In particular we show that, although small for inclusive 
    quantities,
    weak corrections can be sizeable for differential distribution.
    }
\end{center}

\newpage
\setcounter{page}{1}
\section{Introduction}
Top-quark physics plays an important r\^ole at the Tevatron and will be
an equally important topic at the upcoming LHC. In view of the large
production rate, amounting to $\mathcal{O}(10^8)$ top-quark pairs for an
integrated luminosity of 200 ${\rm fb}^{-1}$, precise and direct
measurements will be possible, which require a similarly 
detailed theoretical understanding of these reactions. 
Both single top-quark production as well as top-quark
pair production have been studied extensively in the past. The
differential cross section for top-quark pair production is known to 
next-to-leading order (NLO) accuracy in quantum chromodynamics (QCD) 
\cite{Nason:1988xz,Nason:1989zy,Beenakker:1989bq,Beenakker:1991ma,%
Bernreuther:2001rq}.
In addition, the resummation of logarithmic enhanced contributions has
been studied in detail in 
\Refs{Laenen:1992af,Kidonakis:1995wz,Berger:1996ad,Catani:1996yz,%
Berger:1998gz,Cacciari:2003fi}.
Recently also the spin correlations between top-quark and antitop-quark
were calculated at NLO accuracy in QCD 
\cite{Bernreuther:2001rq,Bernreuther:2004jv}.

Although formally suppressed through the small coupling, weak
corrections can also be significant due to the presence of large
Sudakov logarithms (see e.g. \Refs{Kuhn:1999nn,Kuhn:2001hz} and 
references therein), which were also studied in the context of $\gamma$-
and $Z$-production at hadron colliders 
\cite{Kuhn:2004em,Kuhn:2005az,Kuhn:2005gv}. The origin of these large
logarithms is easily understood: at high scale the massive gauge
bosons $W$ and $Z$ behave essentially like massless bosons. Collinear
and soft phenomena thus lead to large negative corrections. In strictly
massless theories like QED or QCD these
contributions are cancelled through similar positive terms from the real
corrections. This cancellation does not take place in the case of weak
interaction because the real and virtual emission leads to different 
experimental signatures. 
Given that top-quark pair production at high scale is an ideal tool to
search for new physics it is clear that the precise knowledge of the
weak corrections in this region is of paramount importance.
In \Ref{Beenakker:1993yr} electroweak corrections to top-quark pair
production in hadronic collisions were
investigated for the first time. More precisely, the partonic sub-processes $q\bar q \to t \bar t $ and
$gg \to t \bar t $ were studied. In a subsequent study
\cite{Kao:1999kj}  parity violating 
asymmetries were analysed in a two Higgs doublet model and the
minimal supersymmetric standard model.  

\begin{figure}[htbp]
  \begin{center}
    \leavevmode
        \hspace{-1.3cm}\parbox[t]{4.5cm}{\includegraphics[width=4.5cm]{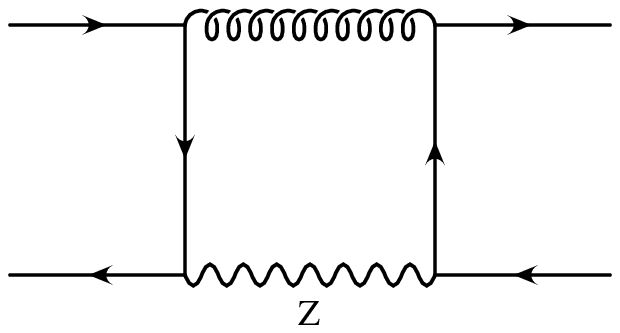}\\
          \centerline{a)}}
        \hspace{1.3cm}\parbox[t]{9.5cm}{
          \vspace*{-2.7cm}

          \parbox[t]{9.5cm}{\large\centerline{%
        $\raisebox{-1.cm}{\includegraphics[width=4.cm]{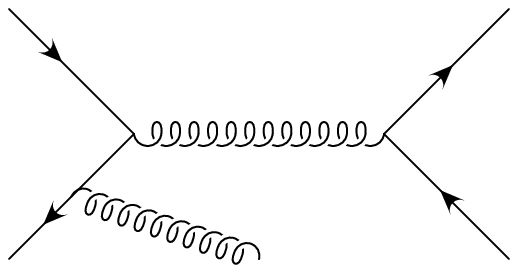}}
        \,\times
        \,\left(\,\,\raisebox{-1.cm}{\includegraphics[width=5.5cm]{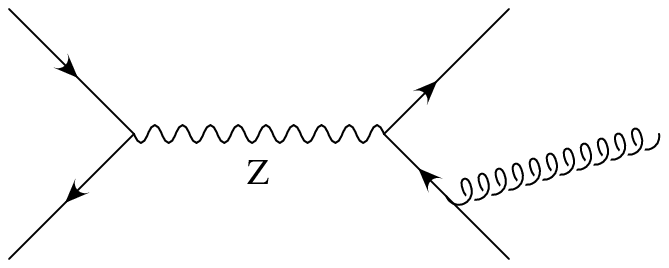}}
        \right)^{\mbox{\Large$\ast$}}$}
      \normalsize\vspace*{0.2cm}
          \centerline{b)}}}\\
        \parbox{6.5cm}{\includegraphics[width=6.5cm]{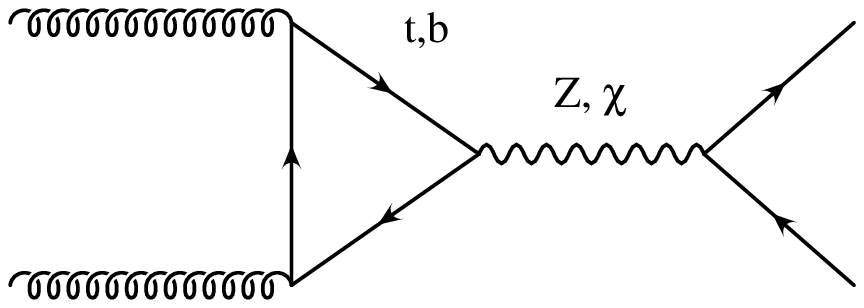}\\
          \centerline{c)}}
        \caption{Pictorial representation of contributions missing in
          \Ref{Beenakker:1993yr}}
        \label{fig:MissingContributions}
  \end{center}
\end{figure}
In the original works some contributions were omitted. 
For the quark--antiquark
initiated process the gluon-$Z$ box contributions 
(Fig.~\ref{fig:MissingContributions} a)  and the corresponding 
real corrections (Fig.~\ref{fig:MissingContributions} b) are missing 
in \Ref{Beenakker:1993yr}. 
They were recently evaluated in \Refs{Kuhn:2005it,Bernreuther:2005is}. 
For the gluon fusion process a class of contributions related to
triangle diagrams (Fig.~\ref{fig:MissingContributions} c) 
are missing in \Ref{Beenakker:1993yr}, as noted also in
\Ref{Moretti:2006nf},
where the calculation of \Ref{Beenakker:1993yr} has been repeated for
the gluon induced top-quark pair production.
In \Ref{Moretti:2006nf} no analytic results are presented. 
In view of the importance of the analysis for upcoming experiments it is 
the purpose of this paper to repeat the original calculation
\cite{Beenakker:1993yr} --- including all missing contributions ---
and present compact analytic results well
suited for the experimental analysis.

The outline of the paper is as follows: In section \ref{sec:virtual} we present
the calculation of the virtual electroweak corrections to top-quark pair
production through gluon fusion. In contrast to the contributions from 
quark--antiquark annihilation there are no real corrections
contributing to the weak corrections. 
The virtual corrections are thus
infrared finite and represent the complete 
weak corrections for this channel. Compact analytic
expressions in terms of scalar one-loop integrals are given in section 
\ref{sec:virtual} and in the appendix.
In section \ref{sec:NumericalResults} we present numerical results 
for the gluon fusion process at the parton level. Furthermore we 
combine the gluon channel with the quark--antiquark annihilation
process (with the elctroweak corrections taken from
\Ref{Kuhn:2005it}), 
fold them with parton distributions and give 
results for the corrections to the total cross section and to $\pt$- and
$\mtt$- distributions relevant for the Tevatron and the LHC.

\section{Electroweak corrections to gluon fusion}
\label{sec:virtual}
\begin{figure}[!htbp]
  \begin{center}
    \leavevmode
    \includegraphics[width=14cm]{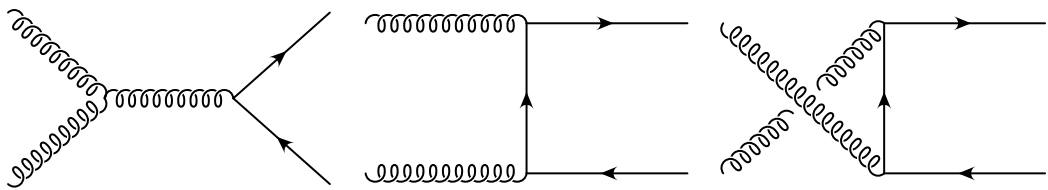}
    \caption{Born diagrams for top-quark pair production via gluon fusion.}
    \label{fig:born}
  \end{center}
\end{figure}
To setup our notation we start with the QCD tree-level contribution.
The three contributions to the amplitude are shown in
Fig.~\ref{fig:born}. Evaluating the Feynman diagrams 
we obtain the well-known leading order 
differential cross section:
\begin{equation}
  {d\sigma_{\rm LO}\over dz} =  
  \sigma_0\*{N^2\*(1+\beta^2\*z^2)-2\over N\*(1-\beta^2\*z^2)^2}\*
  \Big(1-\beta^4\*z^4+2\*\beta^2\*(1-\beta^2)\*(1-z^2)\Big).
  \label{eq:born}
\end{equation}
where $N$ is the number of colours, $\as$ the strong coupling constant
and $\beta$ the velocity of the top-quark in the partonic
centre-of-mass system:
\begin{equation}
  \beta = \sqrt{1-{4\mt^2\over s}}
\end{equation}
($s$ denotes the partonic centre-of-mass energy squared).
The cosine of the scattering angle is denoted by $z$. Here and in what  
follows it is convenient to use the abbreviation
\begin{equation}
  \sigma_0 = {\pi\alpha_s^2\over4}{1\over N^2-1}{\beta\over s}.
  \label{eq:SigmaZero}
\end{equation}
A factor $1/(4\*(N^2-1)^2)$ from averaging over the incoming 
spins and colour is included in the result above.

\begin{figure}[!htbp]
  \begin{center}
    \leavevmode
    \includegraphics[width=9cm]{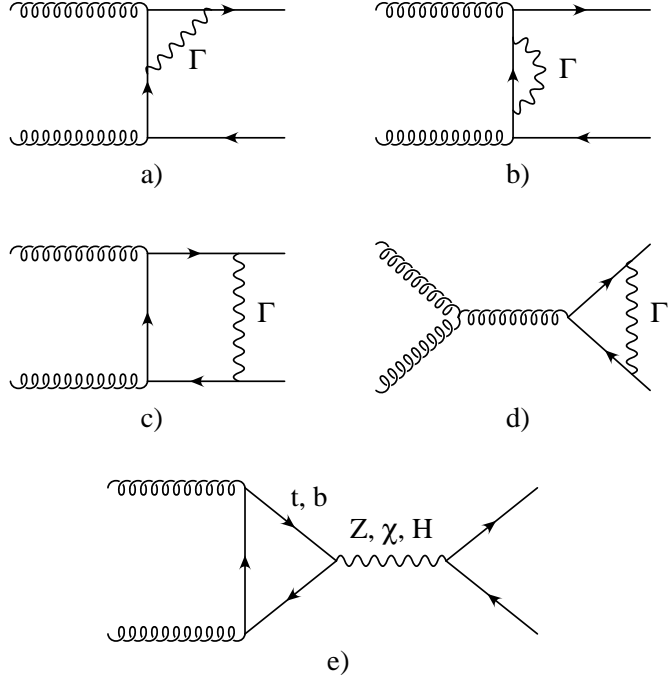}
    \caption{Sample diagrams for the virtual corrections. $\Gamma$
    stands for all contributions from gauge boson, goldstone boson
    and Higgs exchange.}
    \label{fig:loop-diagrams}
  \end{center}
\end{figure}
\begin{figure}[!htbp]
  \begin{center}
    \leavevmode
    \includegraphics[width=10cm]{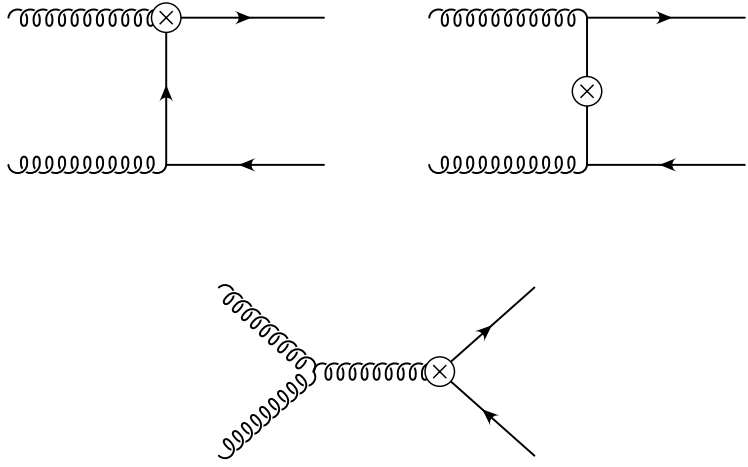}
    \caption{Sample counterterm diagrams.}
    \label{fig:counterterms}
  \end{center}
\end{figure}

For the calculation of the next-to-leading order weak corrections we
use the 't Hooft-Feynman gauge ($R_\xi$-gauge)  with the 
gauge parameters $\xi^i$ set to 1. The longitudinal degrees of freedom
of the massive gauge bosons $Z$ and $W$ are thus represented by the
goldstone fields $\chi$ and $\phi$. Ghost fields do not
contribute at the order under consideration. Sample diagrams are
shown in Fig.~\ref{fig:loop-diagrams}. The Cabibbo--Kobayashi--Maskawa
mixing matrix is set to 1.  
\\
Before presenting the results for the weak corrections, 
let us add a few technical remarks. We
use the Passarino\,--Veltman reduction scheme \cite{PaVe79} to reduce 
analytically the tensor integrals to scalar integrals. 
For these the following convention is used:
\begin{equation}
  X_0 
  = {1\over i\pi^2} \int d^d\ell 
  {(2\pi\mu)^{2\e}\over (\ell^2-m_1^2+i\epsilon)\cdots}.
\end{equation}
For the UV-divergent integrals we define the finite part for the
one-point integrals $A_0$ and the two-point integrals $B_0$ through
\begin{eqnarray}
  A_0(m^2) &=& m^2 \Delta + \overline{A}_0(m^2),\nn\\
  B_0(p^2,m_1^2,m_2^2)    &=&  \Delta+ \overline{B}_0(p^2,m_1^2,m_2^2), 
\end{eqnarray}
with $ \Delta = 1/\e- \gamma + \ln(4\pi)$. 
The renormalisation is performed in the counterterm formalism, where the bare
Lagrangian $\cal L$ is rewritten in terms of renormalised fields and couplings:
\begin{eqnarray}
  {\cal L}(\Psi_0,A_0, m_0,g_0) &=& 
  {\cal L}(Z^{1/2}_\Psi\Psi_R,Z^{1/2}_A A_R, Z_m m_R,Z_g g_R)\nn\\
  &\equiv&  {\cal L}(\Psi_R,A_R, m_R,g_R) 
  + {\cal L}_{ct}(\Psi_R,A_R, m_R,g_R).  
  \label{eq:RenormalizedPerturbationTheory}
\end{eqnarray}
The contribution ${\cal L}(\Psi_R,A_R, m_R,g_R)$ gives just the
ordinary Feynman rules, but with the bare couplings replaced by the
renormalised ones. The complete list of Feynman rules can be found
for example
in \Ref{Denner:1991kt}. The contribution ${\cal L}_{ct}(\Psi_R,A_R,
m_R,g_R)$ in \Eq{eq:RenormalizedPerturbationTheory} 
yields the counterterms,
which render the calculation ultraviolet (UV)-finite. Some of the resulting
diagrams are shown in Fig.~\ref{fig:counterterms}. 
For the present calculation only wave function and mass
renormalisation are needed.
No coupling constant renormalisation has to be performed. 
This is a consequence of the fact, that although the weak corrections
appear as a loop-correction, they are still leading-order in the
electroweak couplings. 
Mass and wave function renormalisation are performed in the on-shell scheme:
\begin{eqnarray}
  m_{t,0} &=&   m_{t} + \delta m_t,\\
  \Psi^{R,L}_{t,0} &=&  \L Z_t^{R,L} \R ^{1/2} \Psi^{R,L}_{t}
  =\L 1 + {1\over 2} \delta Z_t^{R,L} \R  \Psi^{R,L}_{t}.
\end{eqnarray}
The renormalisation constants are thus given in terms of
self-energy corrections $\Sigma$ and their derivatives:
\begin{eqnarray}
  \delta Z_{\rm V} &=& {1\over 2}(\delta Z_t^{L}+\delta Z_t^{R}) = 
  -\Sigma_{\rm V}(p^2=\mt^2)-2\*\mt^2{\partial\over
    \partial p^2}\L\Sigma_{\rm V}+\Sigma_{\rm S}\R\Big|_{p^2=\mt^2}\;, \nn\\
  \delta Z_{\rm A} &=& {1\over 2}(\delta Z_t^{L}-\delta Z_t^{R}) = 
  -\Sigma_{\rm A}(p^2=\mt^2), \nn\\
  {\delta\mt\over \mt} &=& 
  -\Sigma_{\rm V}(p^2=\mt^2)-\Sigma_{\rm S}(p^2=\mt^2).
\end{eqnarray}
The functions $\Sigma_{\rm V,A,S}$ can be found for example in
\Ref{Denner:1991kt}. Here we need only  $\delta Z_{\rm V}$
and ${\delta\mt/ \mt}$. Their explicit form in terms of scalar
integrals --- using the notation of this paper --- reads:
\begin{eqnarray}
  \delta Z_{\rm V} &=&{\alpha\over4\*\pi}\*\Bigg\{\Big[\gvt^2+\gat^2+
  {1\over\mt^2}\*(\gvt^2+\gat^2)\*\Big(\Amzc-\Amtc-\mz^2\*\Bomtmtmzc\Big)\nn\\
  &+&\Big(2\*(\gvt^2+\gat^2)\*\mz^2 +4\*\mt^2\*(\gvt^2-3\*\gat^2)\Big)
  \*\DBomtmtmz\Big]\nn\\
  &+&\*\Vtb\*\Big[{1\over2}+
  {1\over2\*\mt^2}\*\Big(\Amwc-\Ambc\Big)
  -{\mt^2-\mb^2+\mw^2\over2\*\mt^2}\*\Bomtmbmwc \nn\\
  &-&(\mt^2-\mw^2+\mb^2)\*\DBomtmbmw\Big]\nn\\
  &+&\*\cmq\*\Big[
  {1\over 2\*\mt^2}\*\Big(\Amzc-\Amtc-\mz^2\*\Bomtmtmzc\Big)
  \nn\\
  &+&\mz^2\*\DBomtmtmz\Big]\nn\\
  &+&\*\Vtb\*{1\over 4\*\mw^2}\*\Big[
  {\mt^2+\mb^2\over \mt^2}\*\Big(\Amwc-\Ambc\Big) \nn\\
  &-&{(\mt^2-\mb^2+\mw^2)\*(\mt^2+\mb^2)\over\mt^2} \*\Bomtmbmwc \nn\\
  &-&2\*\Big((\mt^2-\mb^2)^2-\mw^2\*(\mt^2+\mb^2)\Big)\*\DBomtmbmw\Big]\nn\\
  &+&\*\hi\*\Big[
  {1\over2\*\mt^2}\*\Big(\Amhc-\Amtc-\mh^2\*\Bomtmtmhc\Big) \nn\\
  &-&(4\*\mt^2-\mh^2)\*\DBomtmtmh\Big]\Bigg\},
\label{eq:Zv}
\end{eqnarray}
\begin{eqnarray}
  {\delta\mt\over \mt} &=&
  -{\alpha\over 4\*\pi}\*\Bigg\{{1\over\mt^2}\Big[(3\*\gat^2-\gvt^2)\*\mt^2+
  (\gvt^2+\gat^2)\Big(\Amtc-\Amzc\Big) \nn \\
  &+& \Big((\gvt^2+\gat^2)\*\mz^2+2\*\mt^2\*(\gvt^2-3\*\gat^2)\Big)
  \*\Bomtmtmzc \Big]\nn\\
  &+&{1\over 4\*s_W^2\*\mt^2}\*\Big[\mt^2 +
  \*\Big(\Ambc-\Amwc\Big)+(\mw^2-\mb^2-\mt^2)\*\Bomtmbmwc \Big]\nn\\
  &+&{1\over 8\*s_W^2\*\mw^2}\*\Big[
  \Amtc-\Amzc +\mz^2\Bomtmtmzc \Big]\nn\\
  &+&{1\over 8\*s_W^2\*\mw^2\*\mt^2}\*\Big[
  (\mt^2+\mb^2)\*\Big(\Ambc-\Amwc\Big)\nn\\
  &-&\Big((\mt^2-\mb^2)^2-\mw^2\*(\mt^2+\mb^2)\Big)\*\Bomtmbmwc \Big]\nn\\
  &+&{1\over 8\*s_W^2\*\mw^2}\*\Big[
  \Amtc-\Amhc+(\mh^2-4\*\mt^2)\*\Bomtmtmhc \Big]\Bigg\},
  \label{eq:Zmt}
\end{eqnarray}
where $\sw$ ($\cw$) denotes the sine (cosine) of the weak mixing angle.
The vector $(\gvt)$ and axial-vector  $(\gat)$ 
couplings of the top-quark to the
$Z$-boson are given in terms of the weak isospin  $T_3^f$
and the electric charge $Q_f$ for a fermion of flavour $f$:
\begin{eqnarray}
  \gvf &=& {1\over 2\sw\cw}( T_3^f - 2 \sw^2 Q_f ),\\
  \gaf &=& {1\over 2 \sw \cw } T_3^f.
\end{eqnarray}
The coupling of the top-quark to the $W$-boson is given by
\begin{equation}
\gw = {1\over2\sqrt{2}\sw}.
\end{equation}
As usual $\alpha$ stands for the fine structure constant, which will be
taken as running coupling evaluated at the scale $2\mt$. The mass
of particle $i$ is denoted by $m_i$.
The abbreviations $B_0^{Z,W,H}$ for the two-point scalar
loop-integrals are defined in the appendix. 
Note that the photonic corrections form a gauge 
independent subset and are not included in Eqs. \ref{eq:Zv},
\ref{eq:Zmt} and in the following discussion.

For the following discussion it is convenient to separate the weak 
corrections into the contribution
from vertex-diagrams ($t$-, $u$-channel, $s$-channel, 
Fig.~\ref{fig:loop-diagrams}~a,d), self-energy-diagrams 
(Fig.~\ref{fig:loop-diagrams}~b) and box-diagrams
(Fig.~\ref{fig:loop-diagrams}~c). The triangle diagrams
in (Fig.~\ref{fig:loop-diagrams}~e) are finite without 
renormalisation and will be studied separately. 
The differential cross section at next-to-leading order is
decomposed as follows:
\begin{eqnarray}
  \label{eq:Decomposition}
  {d\sigma^{\rm NLO}\over dz} = 
  \sum_{i=Z,W,\chi,\phi,H} {d\sigma_i^{\Box}\over dz} 
  + {d\sigma_i^{V}\over dz} 
  + {d\sigma_i^{sV}\over dz} 
  + {d\sigma_i^{{\Sigma}}\over dz} +
  \sum_{i=Z,\chi,H}{d\sigma_i^{{\bigtriangleup}}\over dz}.
\end{eqnarray}

We start with the analytic results for the triangle
diagrams. 
For the Higgs and $(Z + \chi)$-terms we obtain   
\begin{eqnarray}
  {d\sigma^{\bigtriangleup}_{H}\over dz}&=&
  {\alpha\*\over\pi}\*\sigma_0
  \*{\mt^2\over\mw^2\*s_W^2}\*{\beta^2\over1-\beta^2\*z^2}
  \*{1\over s-\mh^2} \nn\\
  &&\*\Big[\mt^2\*(s-4\*\mt^2)\*\Cosmtmtmt+\mb^2\*(s-4\*\mb^2)\*\Cosmbmbmb
  - 2\*(\mt^2+\mb^2)\Big],\nn\\
  && 
  \label{eq:triH}
\end{eqnarray}
\begin{equation}
  {d\sigma^{\bigtriangleup}_{Z+\chi}\over dz} = 
  16\*{\alpha\over \pi}\*\sigma_0
  \*{\gat}\*{\mt^2\over\mz^2\*(1-\beta^2\*z^2)}\*
  \Big(\gat\*\mt^2\*\Cosmtmtmt+g_a^b\*\mb^2\*\Cosmbmbmb\Big),
  \label{eq:triZ}
\end{equation}
A factor $1/(4\*(N^2-1)^2)$ from averaging over the incoming 
spins and colour is again included. The integrals $C_0^{b,t}$ 
are defined in the appendix. 
As a consequence of Furry's theorem only the axial-vector induced terms 
contribute in the case where the $Z$-boson appears in the $s$-channel.
Furthermore, the Landau--Yang theorem 
forbids the decay of an on-shell vector boson into two identical
massless on-shell spin-one bosons. Therefore, the poles from the
propagators of the $Z$-boson and the $\chi$ are cancelled in the \linebreak
$\sigma^{\bigtriangleup}_{Z+\chi}$-term, as evident from \Eq{eq:triZ}.

For the remaining vertex corrections with a gluon in the $s$-channel we obtain
\begin{eqnarray}
  {d\sigma^{sV}_Z\over dz} &=&-2\*{\alpha\over\pi}\*
  \sigma_0\*\genfacs\*\Bigg\{\nn\\
  &&2\*\Big[2\*(\gvt^2+\gat^2)\*\mz^2-s\*\beta^2\*(\gvt^2-3\*\gat^2)\Big]
  \*\Big(\Bosmtmt-\Bomtmtmz\Big)\nn\\
  &+&\Big[4\*(\gvt^2+\gat^2)\*\mz^4+8\*\gat^2\*\mz^2\*s\*\beta^2\nn\\
  &-&s^2\*\beta^2\*(\gvt^2+\gat^2+\beta^2\*(\gvt^2-3\*\gat^2))\Big]
  \*\Cosmtmtmz\nn\\
  &+&s\*\beta^2\*\Big[2\*\mz^2\*(\gvt^2+\gat^2)\nn\\
  &+&s\*(1-\beta^2)\*(\gvt^2-3\*\gat^2)\Big]\*\DBomtmtmz\Bigg\},
\end{eqnarray}
\begin{eqnarray}
  {d\sigma^{sV}_W\over dz} &=&
  -8\*{\alpha\over\pi}\*\sigma_0\*\gw^2\*\genfacs\*\Bigg\{\nn\\
  &&{1\over2}\*(s\*(1+\beta^2)+4\*(\mw^2-\mb^2))\*
  \Big(\Bosmbmb-\Bomtmbmw\Big)\nn\\
  &+&{1\over8}\*\Big(\big(s\*(1+\beta^2)+4\*(\mw^2-\mb^2)\big)^2
  -4\*\beta^2\*s^2\Big)\*\Cosmbmbmw\nn\\
  &-&{1\over4}\*s\*\beta^2\*\Big(s\*(1-\beta^2)+4\*(\mb^2-\mw^2)\Big)\*
  \DBomtmbmw\Bigg\}, \end{eqnarray}
\begin{eqnarray}
  {d\sigma^{sV}_{\chi}\over dz}&=&-8\*{\alpha\over\pi}\*{\sigma_0}
  \*\chifacs\*\genfacschi
  \*\Bigg\{2\*\Big(\Bosmtmt-\Bomtmtmz\Big) \nn\\
  &+&(2\*\mz^2+s\*\beta^2)\*\Cosmtmtmz+s\*\beta^2\*\DBomtmtmz\Bigg\},
\end{eqnarray}
\begin{eqnarray}
  {d\sigma^{sV}_{\phi}\over dz} &=&-8\*{\alpha\over\pi}\*\sigma_0
  \*{\gw^2\over\mw^2}\*\genfacs\*\Bigg\{
  {1\over16}\*\Big(16\*\mb^4+\mb^2\*(8\*s\*\beta^2-16\*\mw^2)\nn\\
  &-&4\*\mw^2\*s\*(1-\beta^2)-s^2\*(1-\beta^4)\Big)\*
  \Big(\Bomtmbmw-\Bosmbmb\Big)\nn\\
  &+&{1\over64}\*\Big[64\*\mb^6-16\*\mb^4\*(8\*\mw^2+s\*(1-\beta^2))\nn\\
  &+&16\*\mw^4\*s\*(1-\beta^2)+4\*\mb^2\*(16\*\mw^4-s^2\*(1-\beta^2)^2)\nn\\
  &+&8\*\mw^2\*s^2\*(1-\beta^4)+s^3\*(1-\beta^2)^3\Big]\*\Cosmbmbmw\nn\\
  &-&{1\over32}\*\Big[16\*\mb^4\*s\*\beta^2-8\*\mb^2
  \*(2\*\mw^2\*s\*\beta^2+s^2\*\beta^2\*(1-\beta^2))\nn\\
  &-&4\*\mw^2\*s^2\*\beta^2\*(1-\beta^2)+s^3\*\beta^2\*(1-\beta^2)^2\Big]
  \*\DBomtmbmw\Bigg\},
\end{eqnarray}
\begin{eqnarray}
  {d\sigma^{sV}_H\over dz} &=& 
  4\*{\alpha\over\pi}\*\sigma_0\*\gw^2\*\genfacs\*{\mt^2\over\mw^2}\*
  \Bigg\{\nn\\
  &&2\*(s\*\beta^2+\mh^2)\*\Big(\Bomtmtmh-\Bosmtmt\Big)\nn\\
  &+&\Big(s^2\*\beta^2\*(1-\beta^2)-3\*\mh^2\*s\*\beta^2-2\*\mh^4\Big)
  \*\Cosmtmtmh\nn\\
  &+&s\*\beta^2\*(s\*(1-\beta^2)-\mh^2)\*\DBomtmtmh\Bigg\}.
\end{eqnarray}
The remaining contributions to the differential cross section are listed
in appendix \ref{sec:results}.

Before showing concrete results for hadron colliders we first discuss  
several checks of our result. We performed two independent calculations
yielding also two independent numerical computer codes. We checked
that we obtain the correct structure for the UV singularities yielding a
finite result after renormalisation. In our notation this corresponds
to a stringent test of the coefficients of the $A_0$- and $B_0$-integrals.  
The behaviour of the Higgs corrections
close to threshold and for light Higgs bosons is well understood 
(see for example \Ref{Jezabek:1993eh} and 
references therein). This allows to test the Higgs 
contributions for a very light Higgs near threshold. 
For a $t\tb$-system produced through a vector current 
the Higgs correction is given by the factor
$(1+H_{\rm thr}(r))$ with 
\begin{eqnarray}
  H_{{\rm thr}}(r) &=&{2\*\kappa\over\pi}\*\Bigg\{
  -{1\over12}\*\Bigg[-12+4\*r+(-12+9\*r-2\*r^2)\ln\L r\R\nn\\
  &+& {2\over r}\*(-6+5\*r-2\*r^2)\*\sqrt{r\*(4-r)}
  \*\arccos\L{\sqrt{r}\over2}\R\Bigg] \nn\\
  &-&{\pi\over\sqrt{r}} + {\pi\over2\*\beta}\*
  \arctan\L{2\*\beta\over \sqrt{r}}\R\Bigg\}
  \label{eq:higgstest}
\end{eqnarray}
and
\begin{equation}
  r = {\mh^2\over\mt^2}, 
  \quad \kappa = {\alpha\over4}\*{\mt^2\over\sw^2\*\mw^2}.
\end{equation}
\begin{figure}[!htbp]
  \begin{center}
    \leavevmode
    \includegraphics[width=12cm]{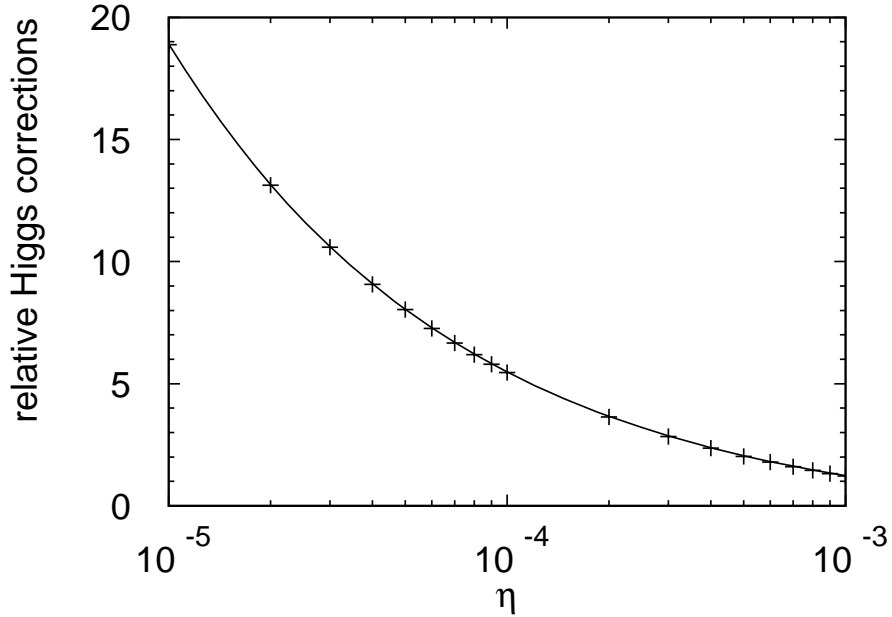}
    \caption{Comparison of the full corrections from Higgs boson
    exchange with the corrections based on \Eq{eq:higgstest} 
    ($\mh = 0.1\mbox{ MeV}$).}
    \label{fig:higgstest}
  \end{center}
\end{figure}%
In Fig.~\ref{fig:higgstest} we show the numerical result for $\mh =
0.1\mbox{ MeV}$ as function of the variable $\eta$ defined through
\begin{equation}
  \eta = {s\over4\mt^2}-1.
\end{equation}
The line shows the Higgs contribution according to \Eq{eq:higgstest}.
The crosses show the result from the full Higgs boson induced
correction. The enhancement proportional to $1/\beta$ is well recovered by
the calculation. The numerical agreement is between four
($\eta = 10^{-5}$) and two digits ($\eta = 10^{-3}$). 
As we shall see later the enhancement close to threshold for a light
Higgs can still be observed even for a Higgs mass of 120~GeV although
reduced to 2\%. 
A similar test has been performed for a light $Z$-boson.
Again we find perfect agreement. 
In addition we compared analytically the results for the 
$t$- and $u$-channel vertices, self-energies and Higgs triangle
diagrams with those from \Ref{Beenakker:1993yr}. We find
complete agreement, after the correction of some typos in 
\Ref{Beenakker:1993yr}.
Using the same input parameters we 
also compared the plots shown in \Ref{Beenakker:1993yr} and found agreement. 
Finally we compared numerically with the results of 
\Ref{Bernreuther:2006xy} and found perfect agreement. 
However, we are in disagreement with the results published recently
in  \Ref{Moretti:2006nf}. In particular for the $\pt$ and $\mtt$ 
distribution we find negative corrections close to threshold while the 
corrections are positive in \Ref{Moretti:2006nf} (Fig. 1). Our
findings are also confirmed in \Ref{Bernreuther:2006xy}. 
\section{Numerical results}
\label{sec:NumericalResults}
\begin{figure}[!htbp]
  \begin{center}
    \leavevmode
    \includegraphics[width=12cm]{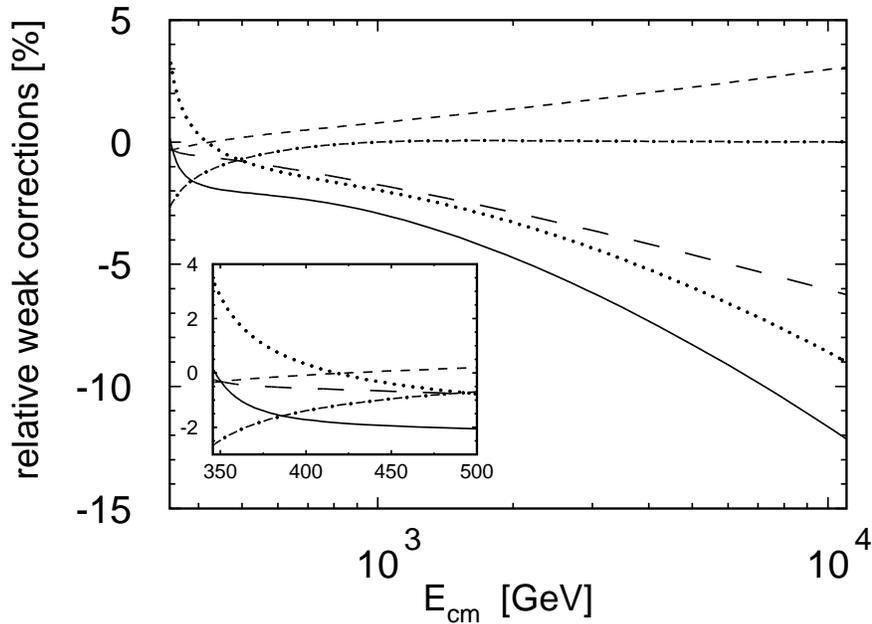}
    \caption{Different contributions to the electroweak corrections:
    Vertices (long-dashed), self-energies (dashed), boxes (dotted),
    triangles (dash-dotted). The sum is shown as full line.}
    \label{fig:singcontri}
  \end{center}
\end{figure}%
In this section we present numerical results for the gluon fusion process at
order $\alpha_s^2\alpha$. 
We use the following coupling constants:
\begin{displaymath}
  \alpha(2\mt) = {1\over 126.3},\quad
  \as = 0.1,\quad
  \sw^2 = 0.231,
\end{displaymath}
and, if not stated otherwise, the following masses:
\begin{displaymath}
  \mz = 91.1876 \GeV,\quad
  \mw = 80.425  \GeV,\quad
  \mh = 120\GeV,\, 
\end{displaymath}
\begin{displaymath}
  \mb = 4.82\GeV,\quad
  \mt = 172.7 \GeV.
\end{displaymath}
The parameters $\sw$, $\mz$ and $\mw$ are, of course,
interdependent. Nevertheless, we expect that, using the $\MSbar$ value for
the weak mixing angle, some of (uncalculated) higher-order
corrections are included and, therefore, a better phenomenological
description is achieved.
The difference between using the $\MSbar$ and the on-shell input respectively,
is formally of higher-order.

\begin{figure}[!htbp]
  \begin{center}
    \leavevmode
    \includegraphics[width=12cm]{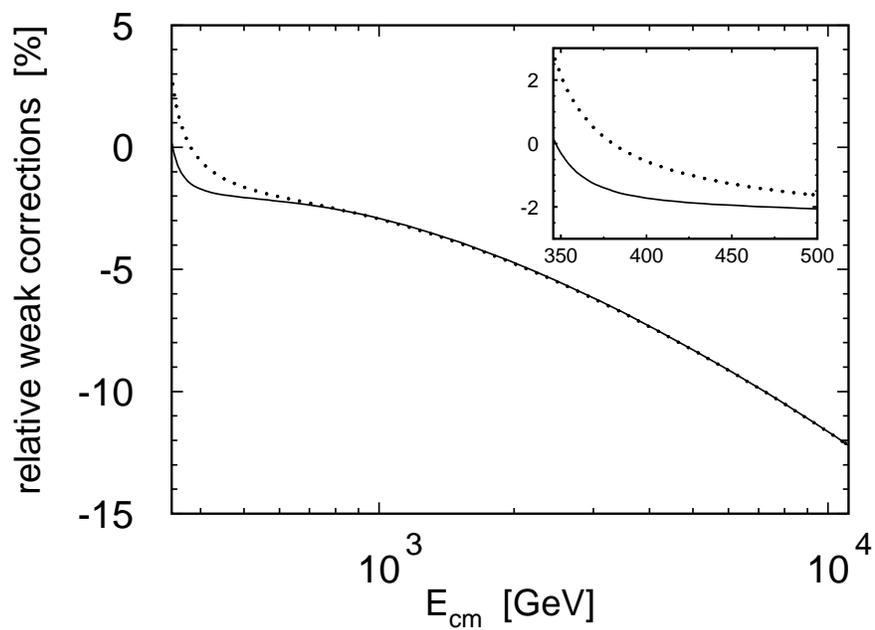}
    \caption{Comparison between the electroweak corrections:
      with the $Z$ and $\chi$ triangle diagrams included (full) 
      and without these contributions (dashed).}
    \label{fig:trivsall}
  \end{center}
\end{figure}

In Fig.~\ref{fig:singcontri} the separate corrections as defined in 
\Eq{eq:Decomposition}
are shown at the parton level. We normalise the result to the
leading-order $gg\to t\bar t$ process.
 The sum of the different contributions
is shown as a solid line in Fig.~\ref{fig:singcontri}.
For moderate partonic energies the corrections are of the order of a few
percent as one might have expected for a weak correction.
Near threshold the cross section is dominated by the triangle-
and the box-diagrams. Both are of the same order of
magnitude, but opposite in sign, leading to a significant cancellation. 
It is worth noting at this
point that the inclusion of the ($Z+\chi$)-triangle diagrams --- neglected in
\Ref{Beenakker:1993yr} --- decreases the result by about 2\% in
the threshold region ($\mh = $ 120~GeV). 
The ($Z+\chi$)-term dominates the triangle contributions. 
In Fig.~\ref{fig:trivsall} we illustrate the effect of $\sigma^{\bigtriangleup}_{Z+\chi}$ by comparing
the full result with the result where $\sigma^{\bigtriangleup}_{Z+\chi}$
is neglected. Close to the threshold the aforementioned 2\% difference
is observed. 
The ($Z+\chi$)-contribution accidentally compensates the positive 
contribution from Higgs exchange, which however, 
becomes small about 20 GeV above threshold. 
For energies above 600~GeV the contribution from the ($Z+\chi$)-triangle
becomes negligible.
The behaviour of the box contribution close to the threshold, shown in
Fig.~\ref{fig:singcontri}, is a consequence of the Higgs effects
discussed before. 

\def\Ecm{{E_{\rm cm}}}
On the other hand for a partonic centre-of-mass energy of around
500 GeV the large negative Sudakov logarithms
start to become important and amount to more than 10 percent at a few TeV. 
If we compare the relative size of the weak correction for gluon and
quark--antiquark induced reactions at large energies, 
we find that they are twice
as large for the quark--antiquark process. This can be 
qualitatively understood by just counting the external lines which can
emit $W$- and $Z$-bosons, and observing that the corrections start to be
dominated by Sudakov logarithms. 
 
\begin{figure}[!htbp]
  \begin{center}
    \leavevmode
    \includegraphics[width=12cm]{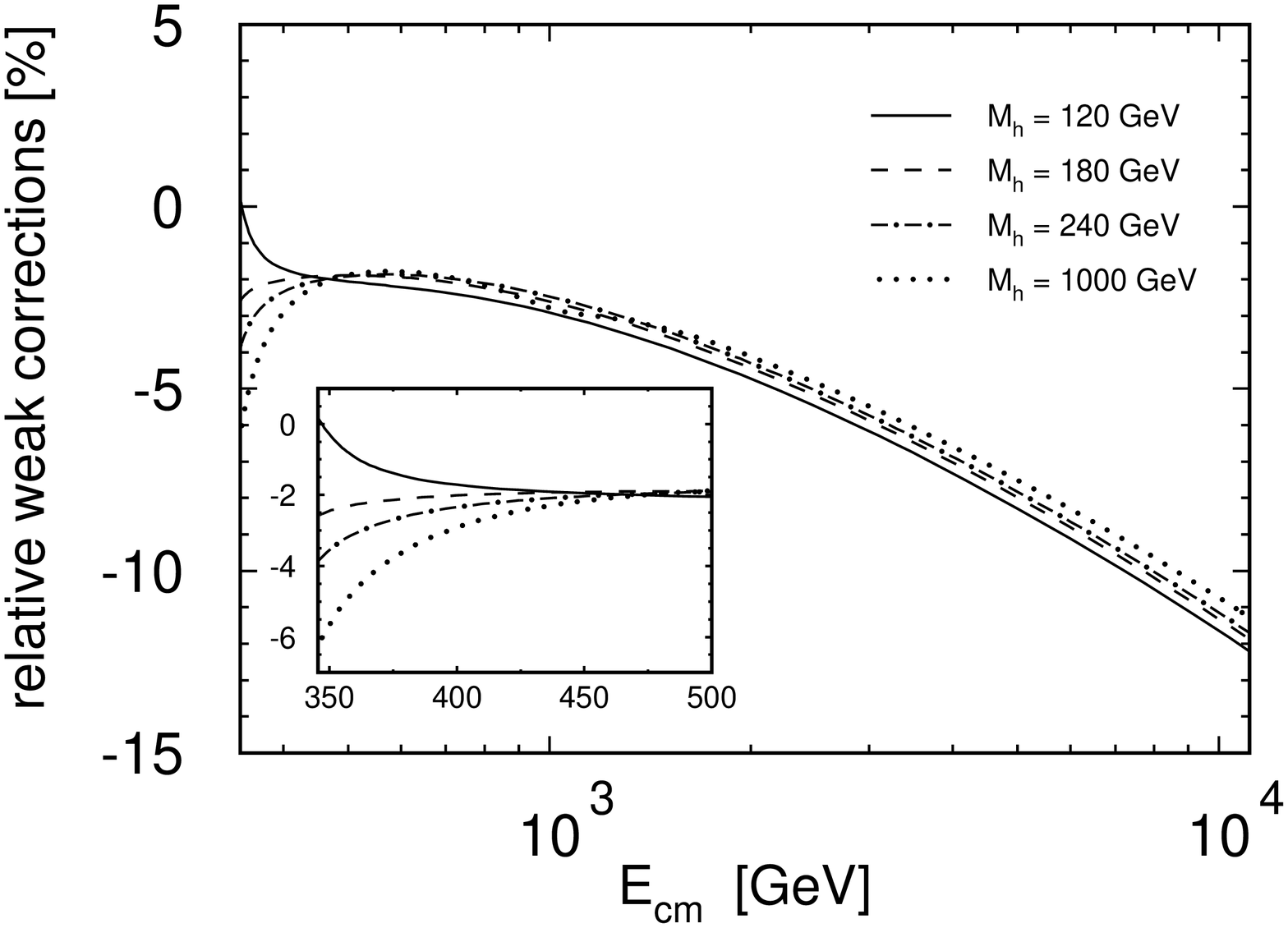}
    \caption{The dependence of the partonic cross section from the Higgs mass:
    $\mh = 120\GeV$ (full line), $\mh = 180\GeV$ (dashed), $\mh =
    240\GeV$ (dashed-dotted) and $\mh = 1000\GeV$ (dotted).}
    \label{fig:higgsdependence}
  \end{center}
\end{figure}%
The dependence on the Higgs mass, i.e. the relative corrections for 
different Higgs masses are
shown in Fig.~\ref{fig:higgsdependence}. For Higgs masses larger than
$2\mt$  we include the width of the Higgs boson in the $s$-channel
propagator. 
\begin{equation}
  {1\over s-\mh^2} \rightarrow {1\over s-\mh^2+i\mh\Gamma_H}.
  \label{eq:bighiggs}
\end{equation}%
The corrections are strongly dependent on $\mh$ with a
variation of nearly 6\% in the threshold region. 
\begin{figure}[!htbp]
  \begin{center}
    \leavevmode
    \includegraphics[width=12cm]{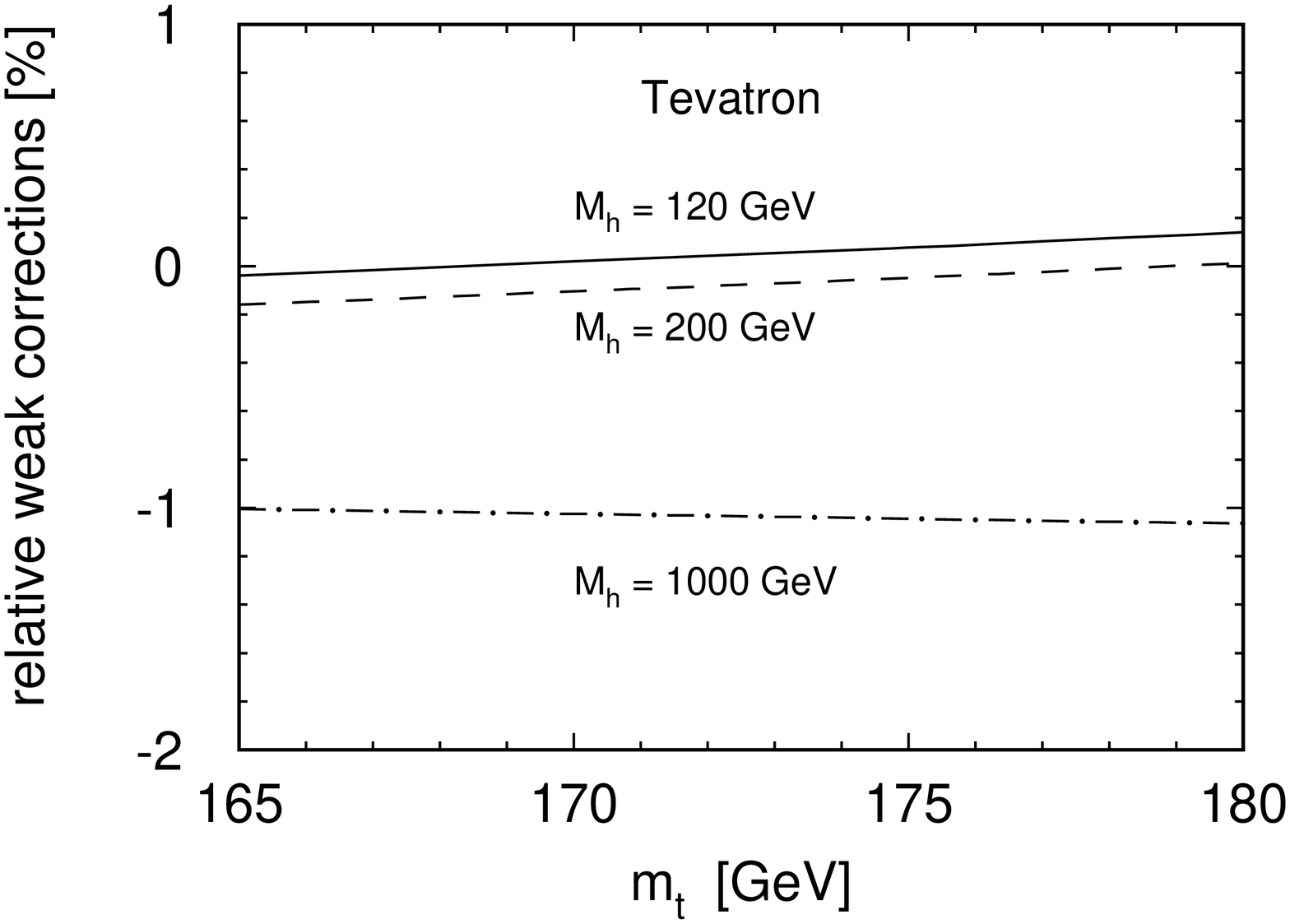}
    \vspace*{-1cm}

    \includegraphics[width=12cm]{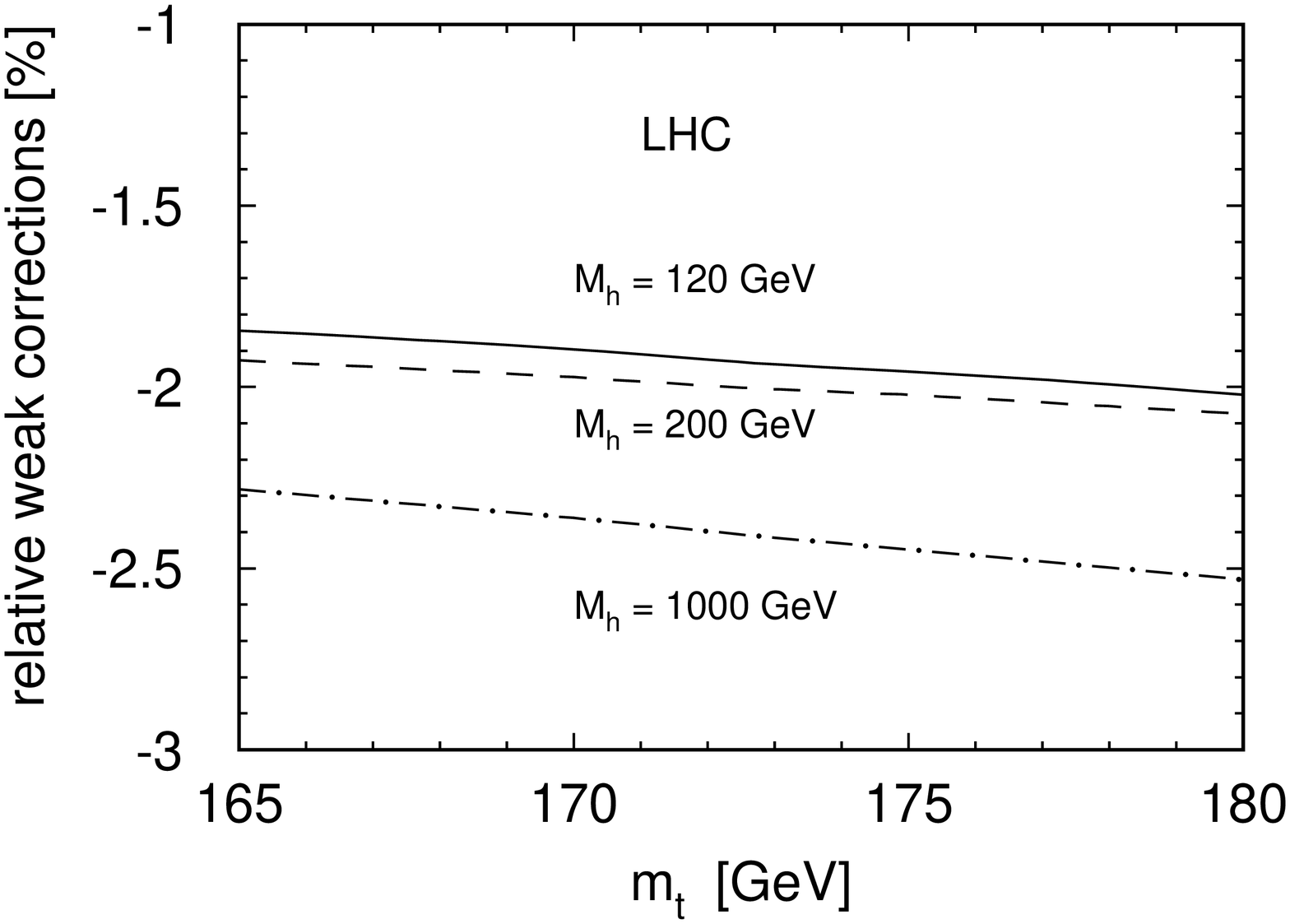}
    \caption{Electroweak corrections to top-quark pair production  
      at the Tevatron (upper figure) and the LHC (lower figure) for 
      three different Higgs masses
      ($\mh = 120\GeV$ (full line), $\mh = 200\GeV$ (dashed), 
      $\mh = 1000\GeV$ (dashed-dotted)).}
    \label{fig:TEVLHC}
  \end{center}
\end{figure}%
Let us now address the effects of the weak corrections on hadronic observables.
We used the parton distribution function CTEQ6L \cite{Kretzer:2003it} 
evaluated at a factorisation scale $\mu_F=2\mt$. With the input 
mentioned before we obtain 
\begin{eqnarray}
  \sigma^{\rm TeV} &=& \sigma_{q\bar{q}}^{\rm TeV} + \sigma_{gg}^{\rm TeV} =
  4.18 \,\, {\rm pb} + 0.19 \,\, {\rm pb} = 4.37 \,\, {\rm pb}\nn
\end{eqnarray}
and
\begin{eqnarray}
  \sigma^{\rm LHC} &=& \sigma_{q\bar{q}}^{\rm LHC} + \sigma_{gg}^{\rm LHC} =
  56 \,\, {\rm pb} + 366 \,\, {\rm pb} = 422 \,\, {\rm pb}\nn
\end{eqnarray}
for the leading-order cross section at the Tevatron respectively LHC.
The leading-order estimate is significantly smaller than the QCD
corrected (NLO + resummation) result of about 6.7~pb for the Tevatron 
\cite{Cacciari:2003fi} and about 794~pb for the LHC
\cite{Bonciani:1998vc}. To some extend the large QCD corrections are of 
universal character and it is plausible that lowest order and electroweak
corrected (total and differential) cross sections will be affected by
similar corrections. Therefore we will, in the following, only present
their relative size. 
In a first step we study the weak corrections to 
the total cross section at the Tevatron and the LHC as a function of
$\mt$ for three different Higgs masses ( Fig.~\ref{fig:TEVLHC}). The
results include both quark--antiquark annihilation and gluon fusion.
As expected the corrections to the inclusive cross section amount to a
few percent only. Most of the top-quark pairs are produced close to
threshold, where the weak corrections at parton level amount to a few
percent only. The relative corrections are essentially given by the
threshold behaviour of the quark--antiquark channel for the Tevatron and the
gluon channel for the LHC. The integrated cross section samples a wide
range in $\mtt$, and the marked Higgs mass dependence of the partonic
cross section close to threshold is washed out. The correction is nearly
independent of the top-quark mass for values of $\mt$ between 165 and 180 GeV.
Given the experimental precision at the Tevatron and the LHC it is
unlikely that the weak corrections can be seen in the total cross section.
Differential distributions in $\pt$ and $\mtt$ are affected however, 
strongly. Indeed, the
corrections to the $\mtt$-distributions can be directly read off from
Fig.~\ref{fig:higgsdependence}, as far as the gluon induced channel is
concerned. Note that for the  quark--antiquark process the situation
is more involved due to the presence of real corrections
\cite{Kuhn:2005it}. In principle it might be possible to establish the
enhancement induced by the top-quark Yukawa coupling for a relatively
light Higgs boson. However, the difference of 6\% between a light ($120$
GeV) and a heavy ($1000$ GeV) Higgs boson could be masked by QCD
uncertainties which are particularly large in the threshold region. 
In contrast, the weak corrections amount to more than ten percent at high
energy when the Sudakov suppression becomes large. Therefore we study
their effect on differential distributions at large momentum transfer.
With the differential distribution in the top-antitop invariant mass 
($\mtt = \sqrt{(\kt+\ktb)^2}$) being a sensitive tool in the search for new
physics, this question is of particular importance.
At large momentum transfer two competing effects must be considered: The
logarithmically increasing Sudakov logarithms, and the increasing
statistical uncertainty.

\begin{figure}[!htbp]
  \begin{center}
    \leavevmode
    \includegraphics[width=12cm]{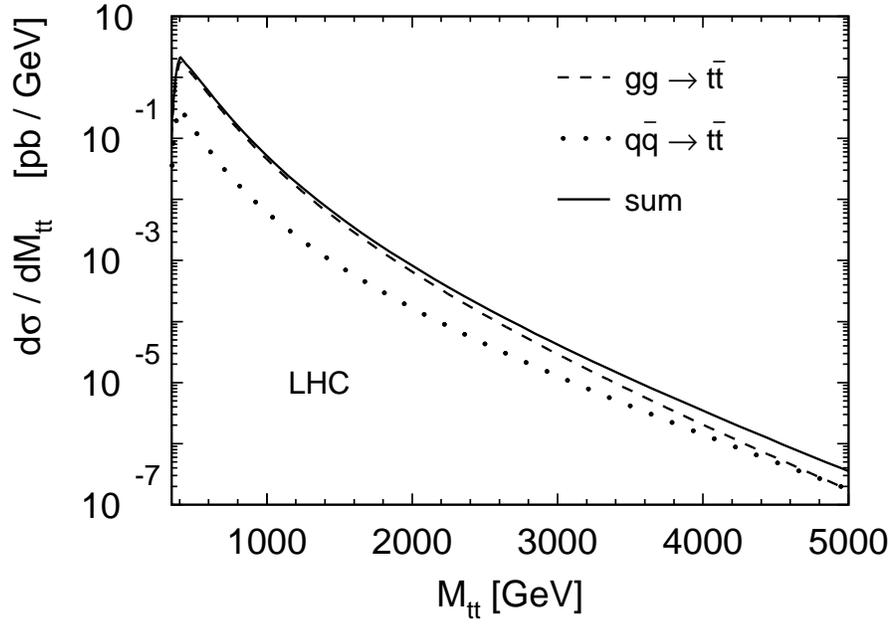}
    \includegraphics[width=12cm]{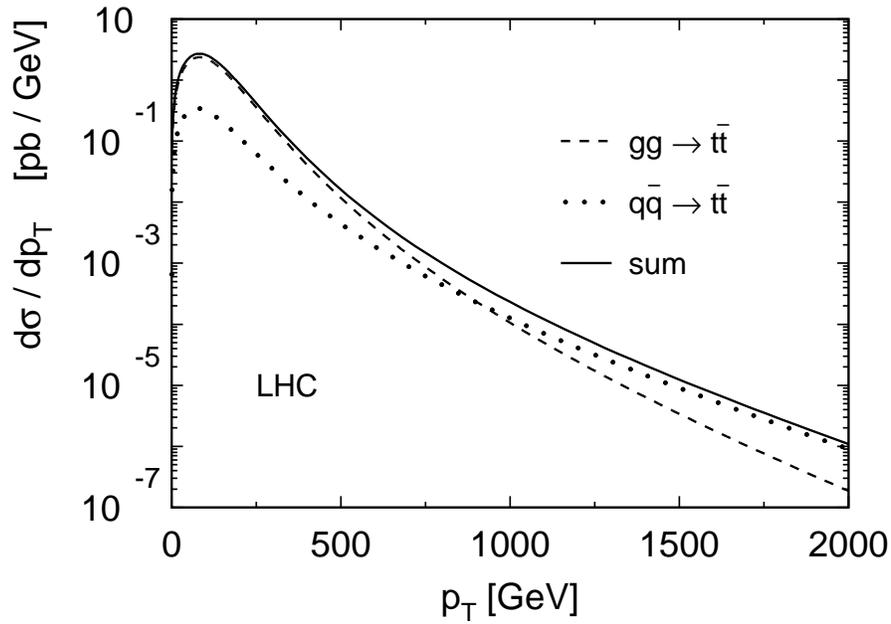}
    \caption{\label{fig:LHCdifplot}%
      Leading-order differential cross section for LHC as a function of
      $\pt$ and $\mtt$. Shown is the sum (full) and the contributions 
      from gluon
      fusion (dashed) and quark--antiquark annihilation (dotted).}
  \end{center}
\end{figure}%
\begin{figure}[!htbp]
  \begin{center}
    \leavevmode
    \includegraphics[width=12cm]{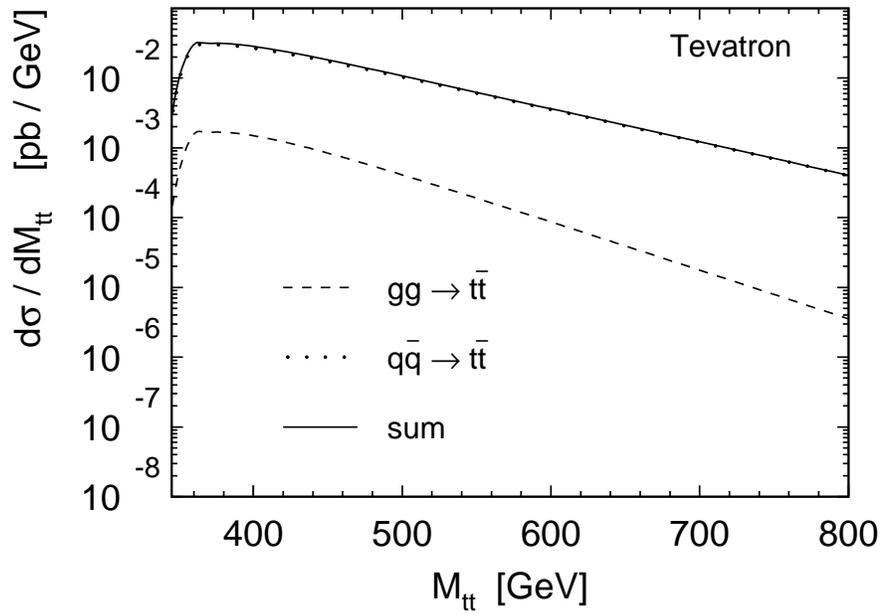}
    \includegraphics[width=12cm]{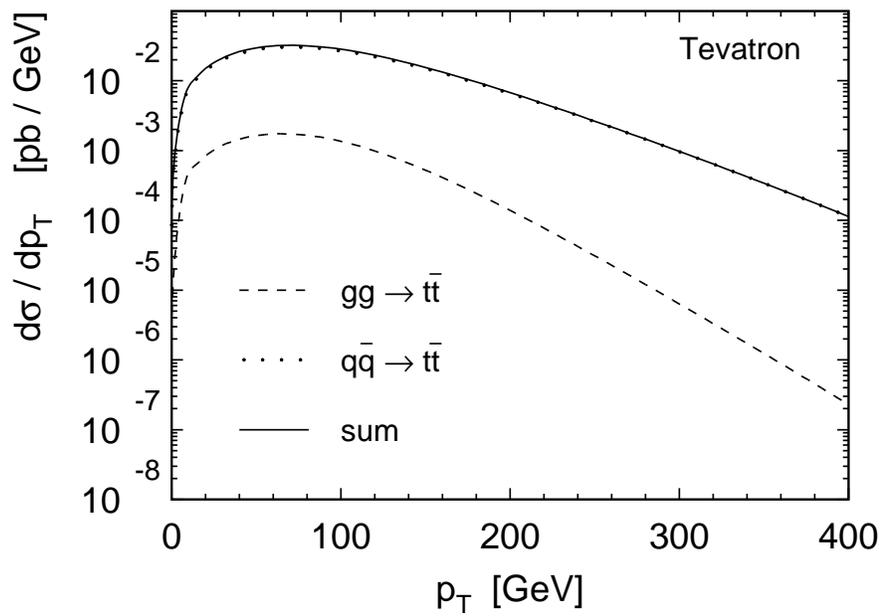} 
    \caption{ \label{fig:TEVdifplot}%
      Leading-order differential cross section for Tevatron as a function of
      $\pt$ and $\mtt$. Shown is the sum (full) and the contributions
      from gluon
      fusion (dashed) and quark--antiquark annihilation (dotted).}
  \end{center}
\end{figure}%
To get a rough idea about the possible statistical sensitivity
we show in Fig.~\ref{fig:LHCdifplot} and 
Fig.~\ref{fig:TEVdifplot} the leading-order differential cross sections
in $\mtt$ respectively $\pt$.
For LHC, about $10^8$ events are expected for an integrated luminosity
of $200\,\, {\rm fb^{-1}}$ and large values of $\mtt$ and $\pt$ will be
accessible. An interesting behaviour of the relative importance of gluon-
versus quark-induced processes is observed for the LHC.
For low $\mtt$ and $\pt$ the gluon channel dominates.
However for $\pt$ larger than 1 TeV the quark--antiquark annihilation
process takes over, as a consequence of the change in relative importance
of quark--antiquark versus gluon luminosities. For the Tevatron only
moderate values of $300$ GeV for $\pt$ and $700$ GeV for $\mtt$ are
accessible at best. Furthermore, the quark--antiquark 
annihilation process dominates through out. 

\begin{figure}[!htbp]
  \begin{center}
    \leavevmode
    \includegraphics[width=12cm]{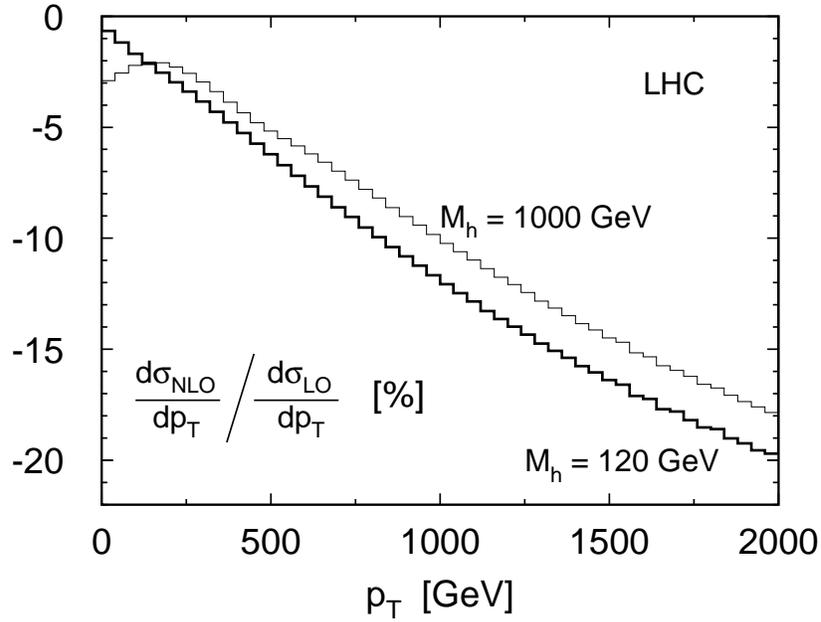}
    \includegraphics[width=12cm]{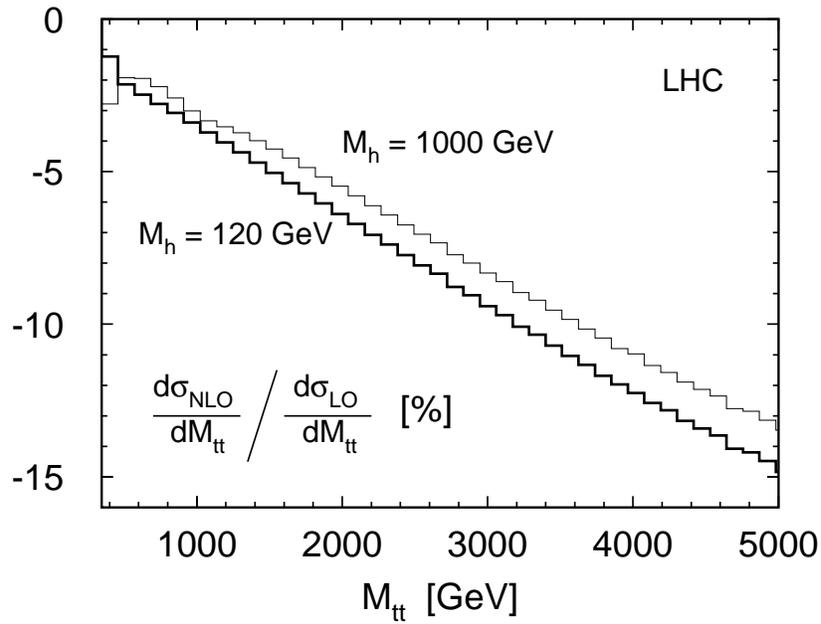}
    \caption{The relative corrections to the $\pt$- and $\mtt$-
    distribution for the LHC for $\mh=120$ GeV (bold
    line) and $\mh=1000$ GeV (thin line).}
    \label{fig:ptmttLHCnlo}
  \end{center}
\end{figure}%
\begin{figure}[!htbp]
  \begin{center}
    \leavevmode
    \includegraphics[width=12cm]{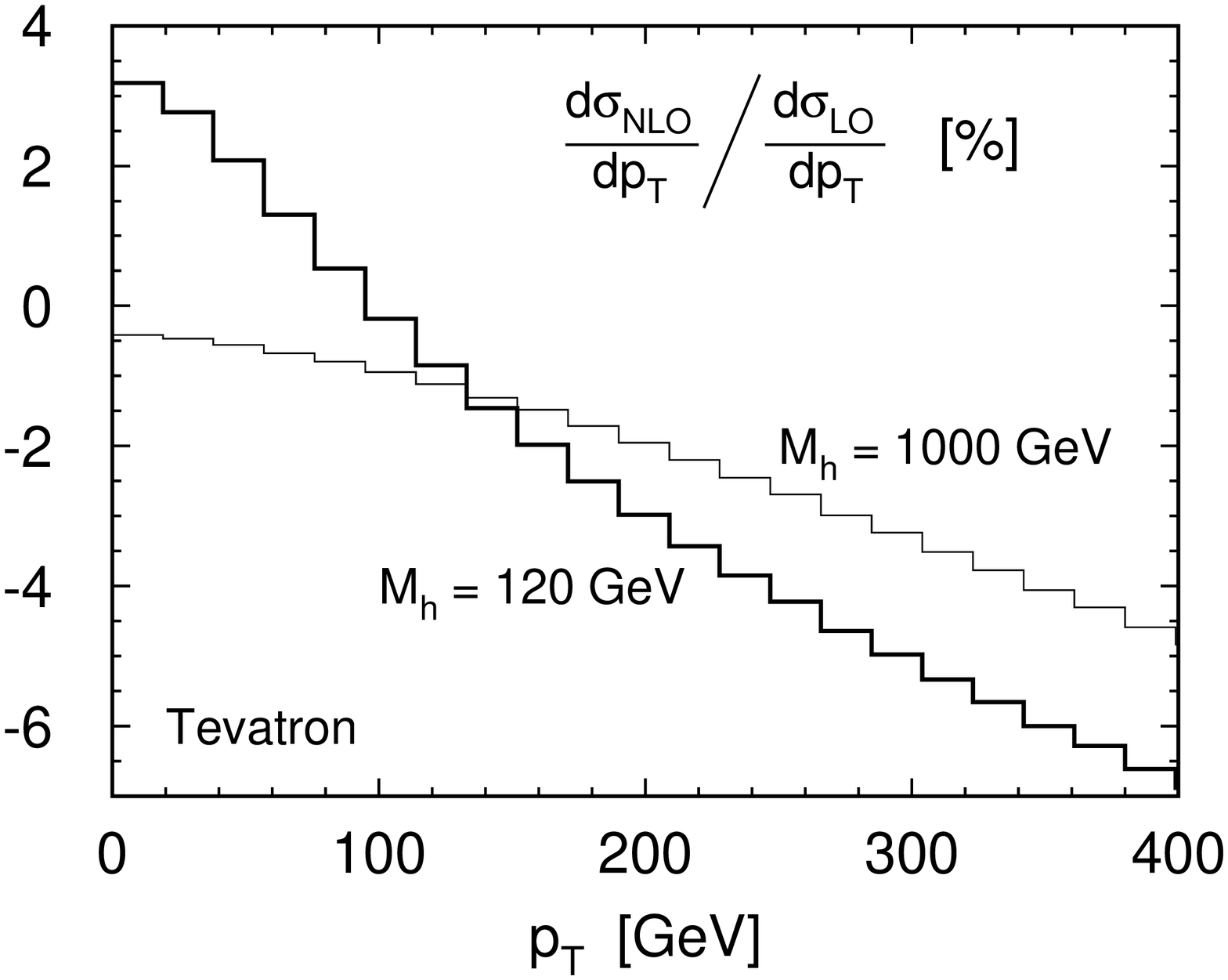}
    \includegraphics[width=12cm]{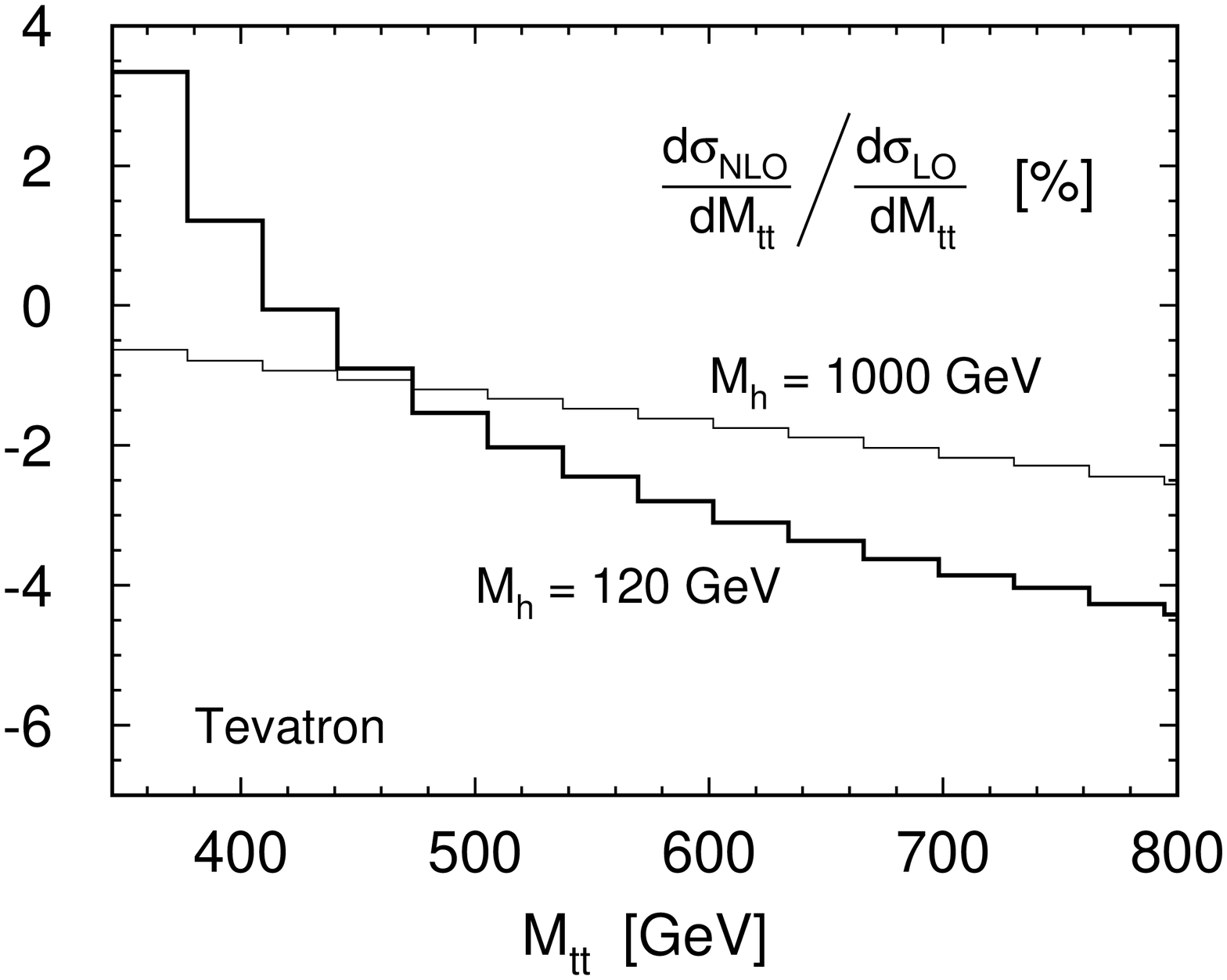}
    \caption{The relative corrections to the $\pt$- and $\mtt$-
    distribution for the Tevatron for  $\mh=120$ GeV (bold
    line) and $\mh=1000$ GeV (thin line).}
    \label{fig:ptmttTEVnlo}
  \end{center}
\end{figure}
\begin{figure}[!htbp]
  \begin{center}
    \leavevmode
    \includegraphics[width=12cm]{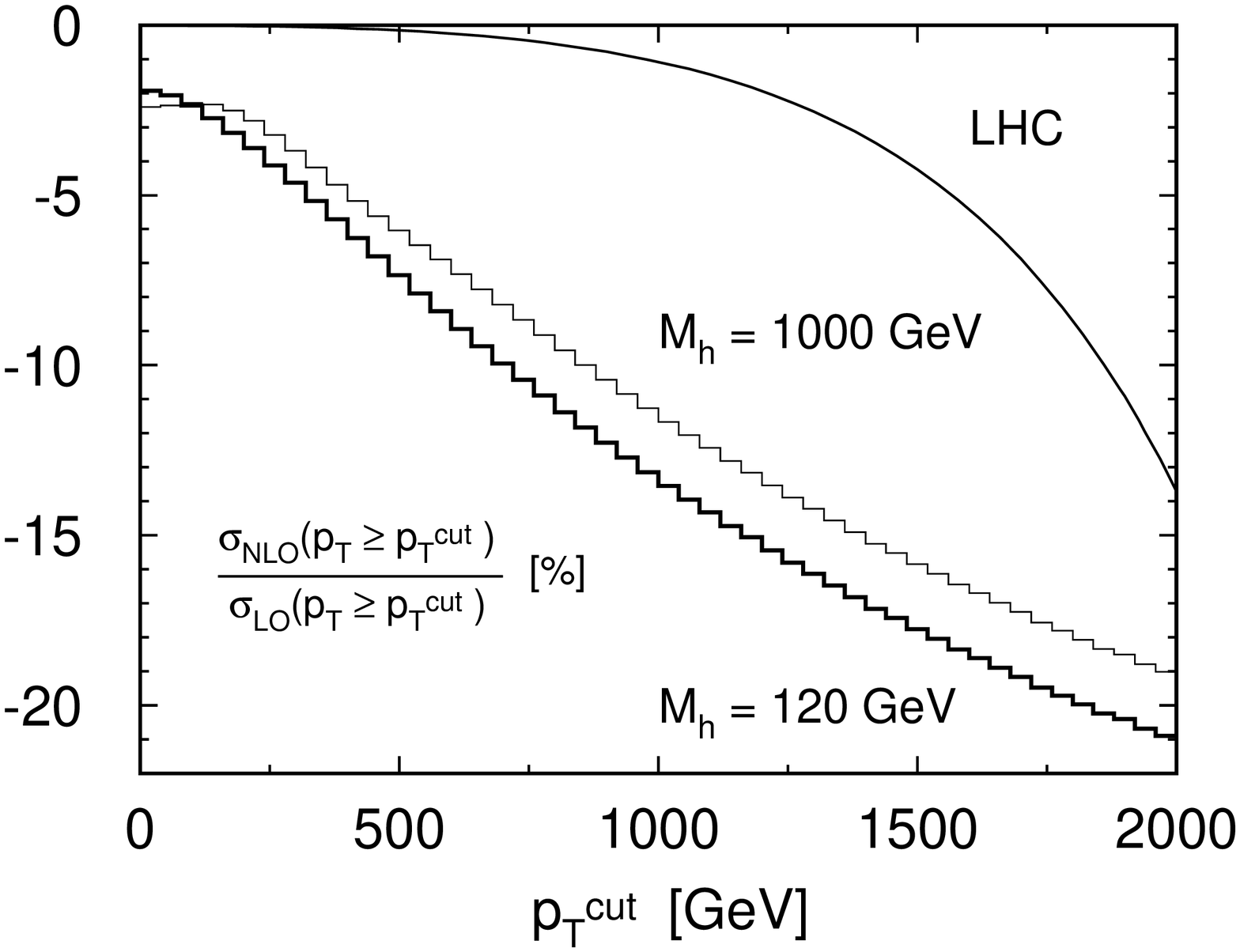}
    \includegraphics[width=12cm]{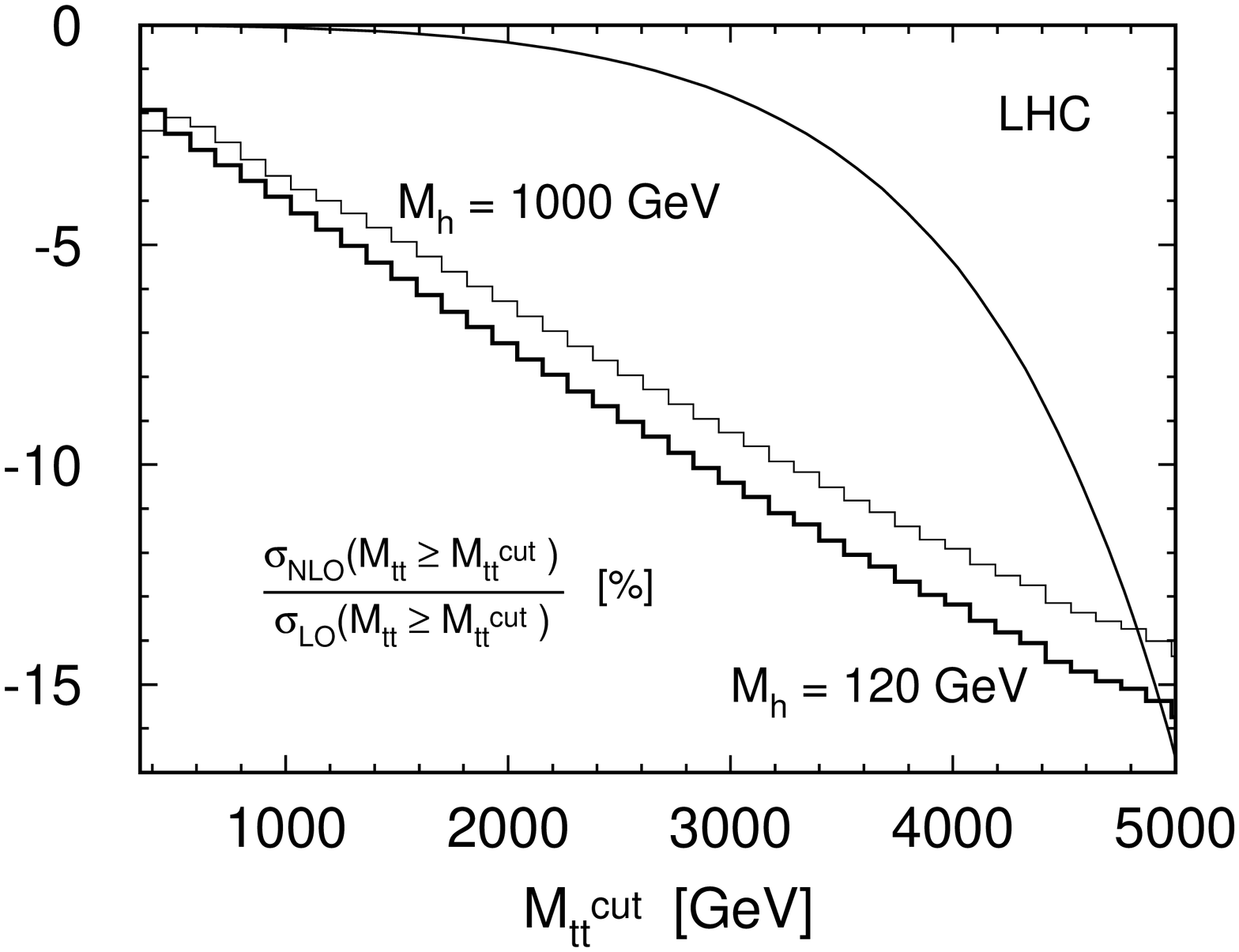}
    \caption{The relative corrections to the $\pt$- and
      $\mtt$-distributions, integrated from a lower value in $\pt$
      respectively $\mtt$ to the kinematic limit, for the LHC and two
      Higgs masses ($\mh=120$ GeV (bold
    line), $\mh=1000$ GeV (thin line)).
    The smooth curve gives an estimate of the corresponding 
    statistical uncertainty for an
    integrated luminosity of $200\,\, {\rm fb^{-1}}$.}
  \label{fig:ptmttLHCnlocut}
  \end{center}
\end{figure}

\begin{figure}[!htbp]
  \begin{center}
    \leavevmode
    \includegraphics[width=12cm]{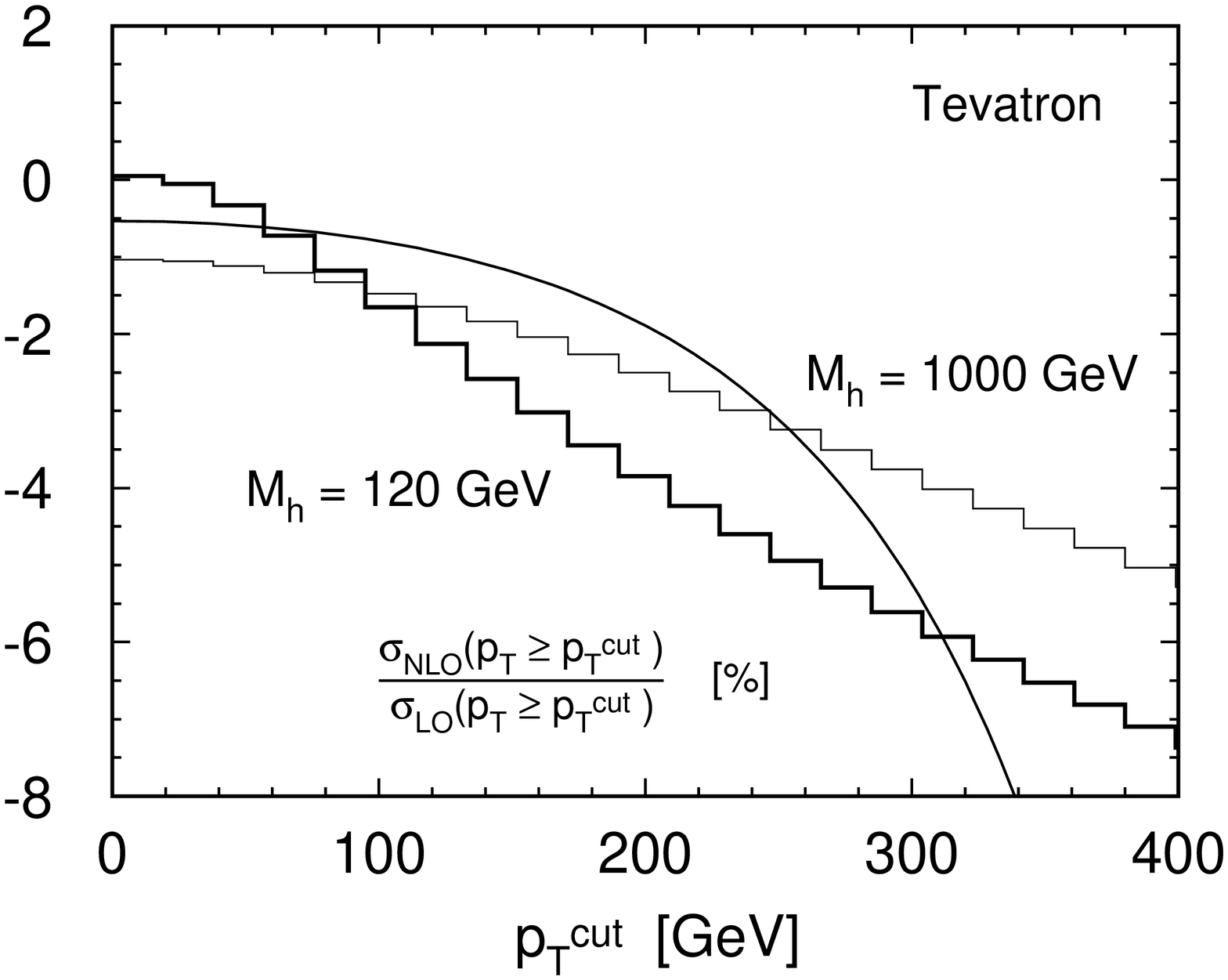}
    \includegraphics[width=12cm]{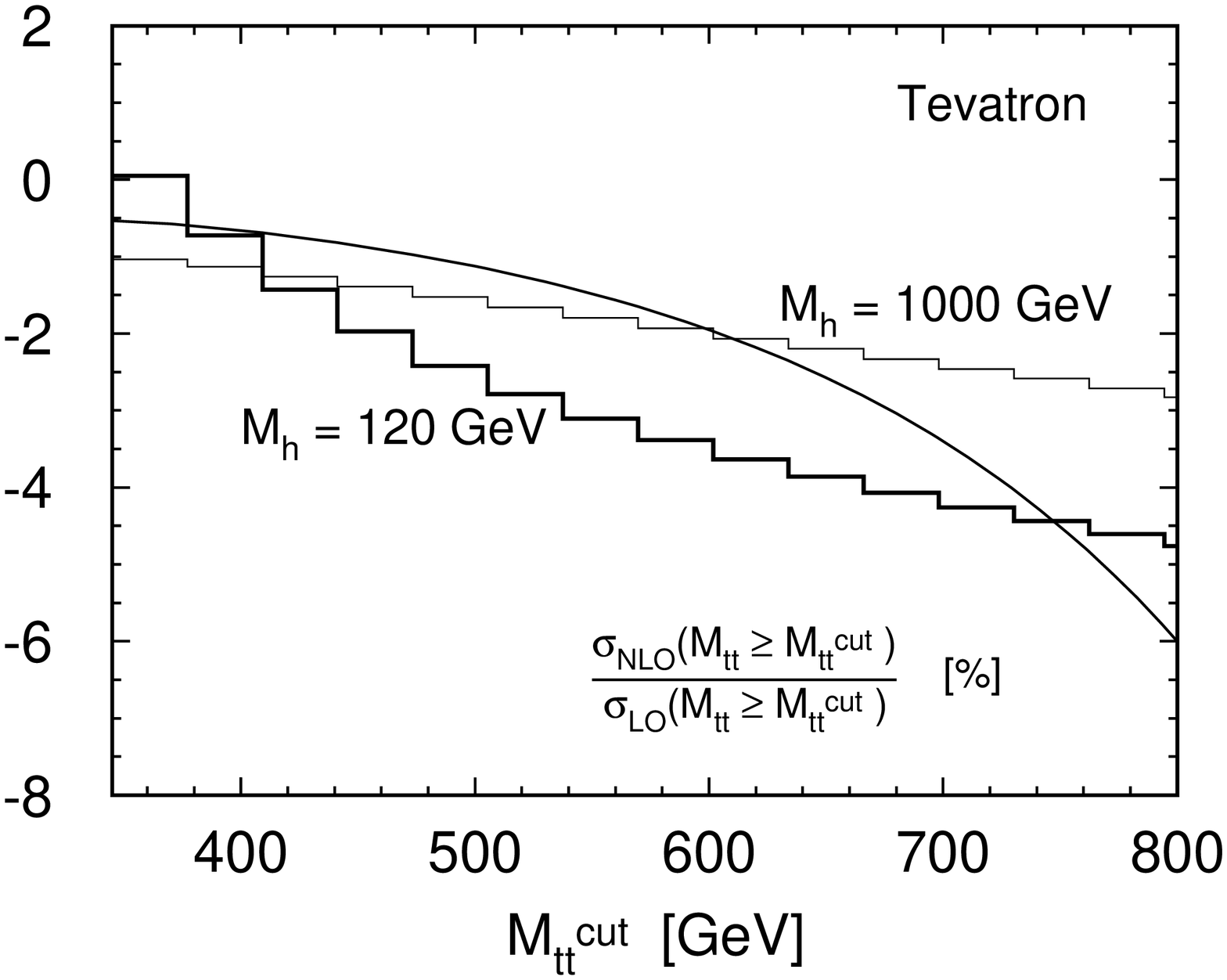}
    \caption{The relative corrections to the $\pt$- and
      $\mtt$-distributions, integrated from a lower value in $\pt$
      respectively $\mtt$ to the kinematic limit, for the Tevatron and two
      Higgs masses ($\mh=120$ GeV (bold line), $\mh=1000$ GeV (thin
      line)). The smooth curve gives an estimate of the corresponding 
      statistical uncertainty for an
      integrated luminosity of $8\,\, {\rm fb^{-1}}$.}
    \label{fig:ptmttTEVnlocut}
  \end{center}
\end{figure}

%
The relative corrections of the differential distributions in $\pt$ and
$\mtt$ are shown in Fig.~\ref{fig:ptmttLHCnlo} and
~\ref{fig:ptmttTEVnlo} for the LHC and Tevatron. For large values of
$\pt$ and $\mtt$, accessible at the LHC, sizeable negative corrections
are predicted, reaching ten up to fifteen percent. In contrast, the
corrections are far smaller at the Tevatron. Taking $\mh = 120\,\, {\rm
  GeV}$, they are $+3\%$ close to threshold and $-5\%$ for $\mtt$ around
$800$ GeV, leading to a distortion of the differential distribution by
$8\%$. It remains to be seen, whether QCD and experimental uncertainties
can be pushed below this level. A rough guess of the statistical
uncertainty expected for LHC and Tevatron can be deduced from 
Figs.~\ref{fig:ptmttLHCnlocut} and Fig.~\ref{fig:ptmttTEVnlocut}. The
estimated number of events with $\mtt \ge \mtt^{\rm cut}$, based on a
luminosity of $200 {\rm fb^{-1}}$ for the LHC ($8 {\rm fb^{-1}}$ for the
Tevatron) is used to estimate the statistical uncertainty and compared with
the relative corrections, evaluated for the same sample. It will be
difficult to observe the effect of the electroweak corrections at the
Tevatron. At the LHC, with the large sample of top quarks, the
statistical precision will match the size of the electroweak Sudakov
logarithms, and eventually of the Higgs enhancement in the threshold
region. 

Before closing this discussion, let us mention
that also the dependence of the
corrections on the bottom-mass was investigated and the full dependence
on the bottom quark mass was kept throughout.
Furthermore the results can also be used to study weak
corrections for $b$-quark pair production. This topic will be discussed
in a subsequent publication. 
For a massless bottom-quark the 
$W$, $\phi$ contributions and the $Z$, $\chi$ triangles differ at most
one percent from the massive case. Hence the effect of the bottom-mass is
negligible at hadron level.

\section{Conclusion}
In this article the complete electroweak corrections
to gluon induced top-quark pair production are calculated. 
In contrast to earlier publications all
contributions of one-loop-order are taken into account and presented in the form of compact
analytic expressions --- well suited to be used in the experimental
analysis. Furthermore the full dependence on the bottom-quark mass is
kept. This allows to calculate also weak corrections for bottom quark 
production using the results presented here.
We have shown that the corrections are negligible for the total cross
section. For differential observables like the $\pt$-distribution or
the distribution in the invariant $t\bar t$-mass the corrections can 
be sizeable. In particular we find corrections up to fifteen  percent in
kinematic regions which are accessible at the LHC. 

{\bf Acknowledgements:} 
We would like to thank W.~Bernreuther, M.~Fücker and Z.-G.~Si for
useful discussions and for a detailed comparison of results prior 
to publication. 
\newpage
\appendix

\section{List of abbreviations}
\label{sec:integrals}
%
%
We define as usual
\begin{eqnarray*}
  &&B_0(p_1^2,\m1^2,\m2^2)={1\over i\pi^2} \int d^d\ell 
  {(2\pi\mu)^{2\e}\over (\ell^2-\m1^2+i\e)
    ((\ell+p_1)^2-\m2^2+i\e)}
  \nn\\ 
\\
  &&C_0(p_1^2,p_2^2,p_1\cdot p_2,
  \m1^2,\m2^2,\m3^2)=\nn\\
  &&{1\over i\pi^2} \int d^d\ell 
  {(2\pi\mu)^{2\e}\over (\ell^2-\m1^2+i\e)
    ((\ell+p_1)^2-\m2^2+i\e)((\ell+p_1+p_2)^2-\m3^2+i\e)}\nn\\
\\
 &&D_0(p_1^2,p_2^2,p_3^2,p_1\cdot p_2,p_1\cdot p_3,p_2\cdot p_3,\m1^2,\m2^2,\m3^2,\m4^2 )=\nn\\
&&{1\over i\pi^2}\int{d^d\ell}{(2\pi\mu)^{2\e}\over
(\ell^2-\m1^2+i\e)((\ell+p_1)^2-\m2^2+i\e)((\ell+p_1+p_2)^2-\m3^2+i\e)}\times\nn\\
&&\qquad\qquad{1\over((\ell+p_1+p_2+p_3)^2-\m4^2+i\e)}
\end{eqnarray*}  
the integrals used in section \ref{sec:virtual} are
  \begin{eqnarray}
    \Bosmtmtc &=& B_0(s,\mt^2,\mt^2)\nn\\
    \Bosmbmbc &=& B_0(s,\mb^2,\mb^2)\nn\\
    \Bomtmtmzc &=& B_0(\mt^2,\mt^2,\mz^2)\nn\\
    \Bomtmbmwc &=& B_0(\mt^2,\mb^2,\mw^2)\nn\\
    \Bomtmtmhc &=& B_0(\mt^2,\mt^2,\mh^2)\nn\\
    \Botmtmzc &=& B_0(-{s\over2}(1+\beta\*z)+\mt^2,\mt^2,\mz^2)\nn\\
    \Botmbmwc &=& B_0(-{s\over2}(1+\beta\*z)+\mt^2,\mb^2,\mw^2)\nn\\
    \Botmtmhc &=& B_0(-{s\over2}(1+\beta\*z)+\mt^2,\mt^2,\mh^2)\nn\\
    \Cosmtmtmt &=& C_0(0,0,{s\over2},\mt^2,\mt^2,\mt^2) \nn\\
    \Cosmbmbmb &=& C_0(0,0,{s\over2},\mb^2,\mb^2,\mb^2)\nn\\
    \Cosmtmtmz &=& C_0(s,\mt^2,-{s\over2},\mt^2,\mt^2,\mz^2)\nn\\
    \Cosmbmbmw &=& C_0(s,\mt^2,-{s\over2},\mb^2,\mb^2,\mw^2)\nn\\
    \Cosmtmtmh &=& C_0(s,\mt^2,-{s\over2},\mt^2,\mt^2,\mh^2)\nn\\
    \Cotmtmtmz &=& C_0(0,\mt^2,-{s\over4}(1+\beta\*z),\mt^2,\mt^2,\mz^2)\nn\\
    \Cotmbmbmw &=& C_0(0,\mt^2,-{s\over4}(1+\beta\*z),\mb^2,\mb^2,\mw^2)\nn\\
    \Cotmtmtmh &=& C_0(0,\mt^2,-{s\over4}(1+\beta\*z),\mt^2,\mt^2,\mh^2)\nn\\
    \Dotz &=& 
    D_0(0,0,\mt^2,{s\over2},-{s\over4}\*(1-\beta\*z),-{s\over4}\*(1+\beta\*z),\mt^2,\mt^2,\mt^2,\mz^2)\nn\\
    \Dotw &=&
    D_0(0,0,\mt^2,{s\over2},-{s\over4}\*(1-\beta\*z),-{s\over4}\*(1+\beta\*z),\mb^2,\mb^2,\mb^2,\mw^2) \nn\\
    \Doth &=&
    D_0(0,0,\mt^2,{s\over2},-{s\over4}\*(1-\beta\*z),-{s\over4}\*(1+\beta\*z),\mt^2,\mt^2,\mt^2,\mh^2)\nn\\
  \end{eqnarray}
To evaluate the Higgs triangle contribution for $\mh > 2\mt$ we give the result for the
corresponding vertex integral:
\begin{equation}
C_0(0,0,{s\over2},m^2,m^2,m^2) = 
{1\over2s}\*\Bigg[\ln\L{1+\beta_m\over1-\beta_m}\R-i\pi\Bigg]^2
\end{equation}
with
\begin{equation}
\beta_m = \sqrt{1-{4m^2\over s}}.
\end{equation}
\section{Analytical results}
\label{sec:results}
The corrections not yet listed in section \ref{sec:virtual} are divided
in self-energy (Fig.~\ref{fig:loop-diagrams}~b), vertex (Fig.~\ref{fig:loop-diagrams}~a)
and box (Fig.~\ref{fig:loop-diagrams}~c) corrections.
Self-energy corrections:
\begin{eqnarray}
   {d\sigma^{\Sigma}_Z\over dz} &=& {\alpha\over8\*\pi}\*\sigma_0\*\genfacself\*\Bigg\{ \nn\\
  &&16\*(\gvt^2+\gat^2)\*{1+\beta^2\*(1-\beta^2)\*(1-3\*z^2)-\beta^4\*z^4\over
  s\*(1-\beta^2)\*(1+\beta^2+2\*\beta\*z)}\*\Big(\Amz-\Amt\Big) \nn\\
  &+&{4\over(1+\beta\*z)^2\*(1+\beta^2+2\*\beta\*z)}\* \nn\\
  && \Big[2\*(\gvt^2+\gat^2)\*{\mz^2\over s}\*(1-z^2)\*\beta^2\*\Big(2+\beta^2-2\*\beta^4+(3\*\beta-2\*\beta^3)\*z+\beta^4\*z^2+\beta^3\*z^3\Big) \nn\\
  &+& \gvt^2\*\Big(2+2\*\beta^2-4\*\beta^4-\beta^6+2\*\beta^8+(4+2\*\beta^2-8\*\beta^4+4\*\beta^6)\*\beta\*z\nn\\
  &+&(-4+7\*\beta^2+\beta^4-3\*\beta^6)\*\beta^2\*z^2-(10-16\*\beta^2+6\*\beta^4)\*\beta^3\*z^3 \nn\\
  &+&(-5+3\*\beta^2+\beta^4)\*\beta^4\*z^4-(4-2\*\beta^2)\*\beta^5\*z^5-\beta^6\*z^6\Big)\nn\\
  &-&\gat^2\*\Big(2+2\*\beta^2-8\*\beta^4-3\*\beta^6+6\*\beta^8+2\*(2-\beta^2-8\*\beta^4+6\*\beta^6)\*\beta\*z \nn\\
  &-&(4-5\*\beta^2-7\*\beta^4+9\*\beta^6)\*\beta^2\*z^2-6\*(1-4\*\beta^2+3\*\beta^4)\*\beta^3\*z^3 \nn\\
  &+&(1-3\*\beta^2+3\*\beta^4)\*\beta^4\*z^4-(4-6\*\beta^2)\*\beta^5\*z^5+\beta^6\*z^6\Big)\Big]\*\Botmtmz \nn\\
  &-&{4\over(1+\beta\*z)^2}\*\Big[{2\over 1-\beta^2}\*(\gvt^2+\gat^2)\*{\mz^2\over s}\* \nn\\
  &&\Big(2+2\*\beta^2-5\*\beta^4+2\*\beta^6+(3-2\*\beta^2)\*\beta^3\*z-3\*(2-3\*\beta^2+\beta^4)\*\beta^2\*z^2\nn\\
  &-&3\*(1-\beta^2)\*\beta^3\*z^3-(2-\beta^2)\*\beta^4\*z^4-z^5\*\beta^5\Big)\nn\\
  &+&(\gvt^2-3\*\gat^2)\*(1-\beta^2)\*(1+\beta^2-2\*\beta^4-(2-3\*\beta^2)\*\beta^2\*z^2-\beta^4\*z^4)\Big]\*\Bomtmtmz\nn\\
  &+&{2\over1+\beta\*z}\*\Big[\big(2\*(\gvt^2+\gat^2)\*\mz^2+s\*(1-\beta^2)\*(\gvt^2-3\*\gat^2)\big)\*\nn\\
  &&\Big(1+(1-\beta^2)\*(2\*\beta^2-\beta\*z-3\*\beta^2\*z^2)-\beta^4\*z^4\Big)\Big]\*\DBomtmtmz\Bigg\}\nn\\
&+&(z\ra-z),
  \end{eqnarray}
\begin{eqnarray}
{d\sigma^{\Sigma}_W\over dz} &=&
{\alpha\over\pi}\*\sigma_0\*\gw^2\*\genfacself\*\Bigg\{ \nn\\
&+&4\*{1+\beta^2\*(1-\beta^2)\*(1-3\*z^2)-\beta^4\*z^4\over
s\*(1-\beta^2)\*(1+\beta^2+2\*\beta\*z)}\*\Big(\Amw-\Amb\Big) \nn\\
&+&{\beta^2\*(1-z^2)\over 2\*s\*(1+\beta^2+2\*\beta\*z)\*(1+\beta\*z)^2}
\*\Big(2+\beta^2-2\*\beta^4+(3-2\*\beta^2)\*\beta\*z+(z+\beta)\*\beta^3\*z^2\Big)\*\nn\\
&&\Big(s\*(1+\beta^2+2\*\beta\*z)+4\*(\mw^2-\mb^2)\Big)\*\Botmbmw \nn\\
&+&{1\over 2\*s\*(1+\beta\*z)^2}\*\Big[
-s\*\beta^2\*(1-z^2)\*\Big(2+\beta^2-2\*\beta^4+(3-2\*\beta^2)\*\beta\*z+(z+\beta)\*\beta^3\*z^2\Big)\nn\\
&-&{4\over1-\beta^2}\*\Big(-2-2\*\beta^2+5\*\beta^4-2\*\beta^6-(3-2\*\beta^2)\*\beta^3\*z+3\*(2-3\*\beta^2+\beta^4)\*\beta^2\*z^2\nn\\
&+&3\*(1-\beta^2)\*\beta^3\*z^3+(2-\beta^2)\*\beta^4\*z^4+z^5\*\beta^5\Big)\*(\mb^2-\mw^2)\Big]\*\Bomtmbmw \nn\\
&+&{1\over4\*(1+\beta\*z)}\*(1+\beta\*(1-\beta^2)\*(2\*\beta-z-3\*\beta\*z^2)-\beta^4\*z^4)\*\nn\\
&&(s\*\beta^2-4\*\mb^2+4\*\mw^2-s)\*\DBomtmbmw \Bigg\}+(z\ra-z),
\end{eqnarray}
\begin{eqnarray}
{d\sigma^{\Sigma}_{\chi}\over dz} &=& 2\*{\alpha\over\pi}\*\sigma_0\*\chifac\*\genfacself\*\Bigg\{ \nn\\
&&{1+(1-\beta^2)\*(\beta^2-3\*\beta^2\*z^2)-\beta^4\*z^4\over2\*(1+\beta^2+2\*\beta\*z)}\*\Big(\Amz-\Amt\Big)\nn\\
&+&{1-\beta^2\over8\*(1+\beta^2+2\*\beta\*z)}\*\Big[2\*\mz^2\*{\beta^2\*(1-z^2)\over(1+\beta\*z)^2}\* \nn\\
&&\Big(2+\beta^2-2\*\beta^4+(3-2\*\beta^2)\*\beta\*z+(z+\beta)\*\beta^3\*z^2\Big)\nn\\
&+&s\*\Big(1+\beta^2-\beta^4-3\*(1-\beta^2)\*\beta^2\*z^2-\beta^4\*z^4\Big)\Big]\*\Botmtmz\nn\\
&-&{\mz^2\over4\*(1+\beta\*z)^2}\*\Big(2+2\*\beta^2-5\*\beta^4+2\*\beta^6+(3-2\*\beta^2)\*\beta^3\*z\nn\\
&-&3\*(2-3\*\beta^2+\beta^4)\*\beta^2\*z^2-3\*(1-\beta^2)\*\beta^3\*z^3-(2-\beta^2)\*\beta^4\*z^4-\beta^5\*z^5\Big)\*\Bomtmtmz\nn\\
&+&{\mz^2\*s\*(1-\beta^2)\over8\*(1+\beta\*z)}\*\Big(1+\beta\*(1-\beta^2)\*(2\*\beta-z-3\*\beta\*z^2)-\beta^4\*z^4\Big)\nn\\
&&\*\DBomtmtmz\Bigg\}+(z\ra-z),
\end{eqnarray}
\begin{eqnarray}
{d\sigma^{\Sigma}_{\phi}\over dz} &=& {\alpha\over\pi}\*\sigma_0\*{\gw^2\over\mw^2}\*\genfacself\*\Bigg\{ \nn\\
&&{(1+\beta^2\*(1-\beta^2)\*(1-3\*z^2)-\beta^4\*z^4)\*(s\*(1-\beta^2)+4\*\mb^2)\over2\*s\*(1-\beta^2)\*(1+\beta^2+2\*\beta\*z)}
\*\Big(\Amw-\Amb\Big) \nn\\
&+&{1\over16\*s\*(1+\beta\*z)^2\*(1+\beta^2+2\*\beta\*z)}\*\Big[\nn\\
&-&\beta^2\*(1-z^2)\*\Big(2+\beta^2-2\*\beta^4+(3-2\*\beta^2)\*\beta\*z+(z+\beta)\*\beta^3\*z^2\Big)\*\nn\\
&&\Big(16\*\mb^2\*(\mb^2-\mw^2)-s^2\*(1-\beta^4+2\*(1-\beta^2)\*\beta\*z)-4\*s\*\mw^2\*(1-\beta^2)\Big) \nn\\
&-&8\*\mb^2\*s\*\Big(-2-2\*\beta^2+4\*\beta^4+\beta^6-2\*\beta^8-2\*(2+\beta^2-4\*\beta^4+2\*\beta^6)\*\beta\*z\nn\\
&+&(4-7\*\beta^2-\beta^4+3\*\beta^6)\*\beta^2\*z^2+2\*(5-8\*\beta^2+3\*\beta^4)\*\beta^3\*z^3+(5-3\*\beta^2-\beta^4)\*\beta^4\*z^4\nn\\
&+&2\*(2-\beta^2)\*\beta^5\*z^5+\beta^6\*z^6\Big)\Big]\*\Botmbmw\nn\\
&-&{1\over16\*s\*(1-\beta^2)\*(1+\beta\*z)^2}\*\nn\\
&&\Big[\beta^2\*s^2\*(1-\beta^2)^2\*(1-z^2)\*\Big(2+\beta^2-2\*\beta^4+(3-2\*\beta^2)\*\beta\*z+(z+\beta)\*\beta^3\*z^2\Big)\nn\\
&-&4\*\Big(-2-2\*\beta^2+5\*\beta^4-2\*\beta^6-(3-2\*\beta^2)\*\beta^3\*z+3\*(2-3\*\beta^2+\beta^4)\*\beta^2\*z^2\nn\\
&+&3\*(1-\beta^2)\*\beta^3\*z^3+(2-\beta^2)\*\beta^4\*z^4+\beta^5\*z^5\Big)\*\Big(s\*\mw^2\*(1-\beta^2)+4\*\mb^2\*(\mw^2-\mb^2)\Big)\nn\\
&+&8\*\mb^2\*s\*(1-\beta^2)^2\*\Big(1+\beta^2-2\*\beta^4-(2-3\*\beta^2)\*\beta^2\*z^2-\beta^4\*z^4\Big)\Big]\*\Bomtmbmw \nn\\
&-&{1\over32\*(1+\beta\*z)}\*\Big(1+\beta\*(1-\beta^2)\*(2\*\beta-z-3\*\beta\*z^2)-\beta^4\*z^4\Big)\*\nn\\
&&\Big(s^2\*(1-\beta^2)^2-4\*s\*(1-\beta^2)\*(2\*\mb^2+\mw^2)\nn\\
&+&16\*\mb^2\*(\mb^2-\mw^2)\Big)\*\DBomtmbmw\Bigg\}+(z\ra-z),
\end{eqnarray}
\begin{eqnarray}
{d\sigma^{\Sigma}_{H}\over dz} &=& {\alpha\over\pi}\*\sigma_0\*{\gw^2\over\mw^2}\*\genfacself\*\Bigg\{ \nn\\
&&{1+\beta^2\*(1-\beta^2)\*(1-3\*z^2)-\beta^4\*z^4\over2\*(1+\beta^2+2\*\beta\*z)}\*\Big(\Amh-\Amt\Big)\nn\\
&-&{(1-\beta^2)\over8\*(1+2\*\beta\*z+\beta^2)\*(1+\beta\*z)^2}\*\nn\\
&&\Big[-2\*\mh^2\*\beta^2\*(1-z^2)\*(2+\beta^2-2\*\beta^4+(3-2\*\beta^2)\*\beta\*z+(z+\beta)\*\beta^3\*z^2)\nn\\
&+&s\*\Big(1+\beta^2-5\*\beta^4-2\*\beta^6+4\*\beta^8+2\*(1-\beta^2-5\*\beta^4+4\*\beta^6)\*\beta\*z \nn\\
&-&(2-2\*\beta^2-5\*\beta^4+6\*\beta^6)\*\beta^2\*z^2-2\*(1-7\*\beta^2+6\*\beta^4)\*\beta^3\*z^3\nn\\
&+&(2-3\*\beta^2+2\*\beta^4)\*\beta^4\*z^4-2\*(1-2\*\beta^2)\*\beta^5\*z^5+\beta^6\*z^6\Big)\Big]\*\Botmtmh\nn\\
&+&{1\over4\*(1+\beta\*z)^2}\*\Big[s\*(1-\beta^2)^2\*(1+\beta^2-2\*\beta^4-(2-3\*\beta^2)\*\beta^2\*z^2-\beta^4\*z^4)\nn\\
&+&\mh^2\*(-2-2\*\beta^2+5\*\beta^4-2\*\beta^6-(3-2\*\beta^2)\*\beta^3\*z+3\*(2-3\*\beta^2+\beta^4)\*\beta^2\*z^2\nn\\
&+&3\*(1-\beta^2)\*\beta^3\*z^3+(2-\beta^2)\*\beta^4\*z^4+\beta^5\*z^5)\Big]\*\Bomtmtmh\nn\\
&-&{s\*(s\*(1-\beta^2)-\mh^2)\*(1-\beta^2)\over8\*(1+\beta\*z)}\*
\Big(1+\beta\*(1-\beta^2)\*(2\*\beta-z-3\*\beta\*z^2)-\beta^4\*z^4\Big)\nn\\
&&\*\DBomtmtmh\Bigg\}+(z\ra-z).\end{eqnarray}
Vertex corrections:
\begin{eqnarray}
{d\sigma^{V}_Z\over dz} &=& {\alpha\over8\*\pi}\*\sigma_0\*\genfacvert\*\Bigg\{4\*(1-\beta^2)\*(\gvt^2+\gat^2) \nn\\
&+&{16\*(\gvt^2+\gat^2)  \over s\*(1-\beta\*z)\*(1-\beta^2)\*(1+2\*\beta\*z+\beta^2)}\*\nn\\
&&\Big(1+2\*\beta^2-\beta^4+(1+4\*\beta^2-3\*\beta^4)\*\beta\*z-2\*(1-\beta^2)\*(2\*\beta^2\*z^2+3\*\beta^3\*z^3)\nn\\
&-&2\*(1+\beta\*z)\*\beta^4\*z^4\Big)\*\Big(\Amt-\Amz\Big) \nn\\
&+&{4\over(1-\beta^2\*z^2)\*(1+2\*\beta\*z+\beta^2)}\*\Big[2\*(\gvt^2+\gat^2)\*{\mz^2\over s}\*
\Big(1-4\*\beta^2-3\*\beta^4+4\*\beta^6 \nn\\
&+&(1-8\*\beta^2+5\*\beta^4)\*\beta\*z+2\*(1+2\*\beta^2-3\*\beta^4)\*\beta^2\*z^2\nn\\
&+&6\*(1-\beta^2)\*\beta^3\*z^3+2\*\beta^6\*z^4+2\*\beta^5\*z^5\Big) \nn\\
&-&\gvt^2\*\Big(2+3\*\beta^2-6\*\beta^4-\beta^6+4\*\beta^8+(3+7\*\beta^2-13\*\beta^4+7\*\beta^6)\*\beta\*z \nn\\
&-&(7-18\*\beta^2+3\*\beta^4+6\*\beta^6)\*\beta^2\*z^2-(14-26\*\beta^2+12\*\beta^4)\*\beta^3\*z^3\nn\\
&-&(10-6\*\beta^2-2\*\beta^4)\*\beta^4\*z^4-(8-4\*\beta^2)\*\beta^5\*z^5-2\*\beta^6\*z^6\Big)\nn\\
&+&\gat^2\*\Big(-2+5\*\beta^2-10\*\beta^4-7\*\beta^6+12\*\beta^8-(3-9\*\beta^2+35\*\beta^4-25\*\beta^6)\*\beta\*z\nn\\
&-&(1-6\*\beta^2-11\*\beta^4+18\*\beta^6)\*\beta^2\*z^2-2\*(1-19\*\beta^2+18\*\beta^4)\*\beta^3\*z^3\nn\\
&+&2\*(1-3\*\beta^2+3\*\beta^4)\*\beta^4\*z^4-4\*(2-3\*\beta^2)\*\beta^5\*z^5+2\*\beta^6\*z^6\Big)\Big]\*\Botmtmz \nn\\
&+&{4\over (1-\beta^2\*z^2)\*(1-\beta^2)}\*\Big[2\*(\gvt^2+\gat^2)\*{\mz^2\over s}\*\Big(1+8\*\beta^2-11\*\beta^4+4\*\beta^6 \nn\\
&+&(1+4\*\beta^2-3\*\beta^4)\*\beta\*z-2\*(5-8\*\beta^2+3\*\beta^4)\*\beta^2\*z^2 \nn\\
&-&6\*(1-\beta^2)\*\beta^3\*z^3-2\*(2-\beta^2)\*\beta^4\*z^4-2\*\beta^5\*z^5 \Big) \nn\\
&+&(1-\beta^2)^2\*\Big(\gvt^2\*(1+\beta^2-4\*\beta^4-(1-\beta^2)\*\beta\*z-2\*(1-3\*\beta^2)\*\beta^2\*z^2-2\*\beta^4\*z^4) \nn\\
&+&\gat^2\*(1-7\*\beta^2+12\*\beta^4-(1-\beta^2)\*\beta\*z+6\*(1-3\*\beta^2)\*\beta^2\*z^2+6\*\beta^4\*z^4)\Big)\Big]\*\Bomtmtmz \nn\\
&+&2\*(\gvt^2+\gat^2)\*s\*(1-\beta^2)^2\*\Cotmtmtmz \nn\\
&-&{4\over1-\beta\*z}\*\Big(((\gvt^2-3\*\gat^2)\*s\*(1-\beta^2)+2\*(\gvt^2+\gat^2)\*\mz^2)\nn\\
&&\*(1+\beta\*(1-\beta^2)\*(2\*\beta-z-3\*\beta\*z^2)-\beta^4\*z^4)\Big)\*\DBomtmtmz\Bigg\}\nn\\
&+&(z\ra-z),\end{eqnarray}
\begin{eqnarray}
{d\sigma^{V}_W\over dz} &=& {\alpha\over\pi}\*\sigma_0\*\gw^2\*\genfacvert\*\Bigg\{(1-\beta^2)\nn\\
&+&{4\over s\*(1-\beta\*z)\*(1-\beta^2)\*(1+2\*\beta\*z+\beta^2) }\*\nn\\
&&\Big(1+2\*\beta^2-\beta^4+(1+4\*\beta^2-3\*\beta^4)\*\beta\*z-2\*(1-\beta^2)\*(2\*\beta^2\*z^2+3\*\beta^3\*z^3)\nn\\
&-&2\*(1+\beta\*z)\*\beta^4\*z^4\Big)\*\Big(\Amb-\Amw\Big)\nn\\
&+&{1\over2\*s\*(1-\beta^2\*z^2)\*(1+2\*\beta\*z+\beta^2)}\*
\Big[4\*(\mw^2-\mb^2)\*\Big(1-4\*\beta^2-3\*\beta^4+4\*\beta^6\nn\\
&+&(1-8\*\beta^2+5\*\beta^4)\*\beta\*z+2\*(1+2\*\beta^2-3\*\beta^4)\*\beta^2\*z^2+6\*(1-\beta^2)\*\beta^3\*z^3\nn\\
&+&2\*\beta^6\*z^4+2\*\beta^5\*z^5\Big)
-s\*(1+2\*\beta\*z+\beta^2)\*\Big(3+3\*\beta^4-4\*\beta^6-(1-8\*\beta^2+5\*\beta^4)\*\beta\*z\nn\\
&-&6\*(1-\beta^4)\*\beta^2\*z^2-6\*(1-\beta^2)\*\beta^3\*z^3-2\*\beta^6\*z^4-2\*\beta^5\*z^5\Big)\Big]\*\Botmbmw\nn\\
&+&{1\over2\*s\*(1-\beta^2)\*(1-\beta^2\*z^2)}\*\Big[4\*(\mw^2-\mb^2)\*\Big(1+8\*\beta^2-11\*\beta^4+4\*\beta^6 \nn\\
&+&(1+4\*\beta^2-3\*\beta^4)\*\beta\*z-2\*(5-8\*\beta^2+3\*\beta^4)\*\beta^2\*z^2 \nn\\
&-&6\*(1-\beta^2)\*\beta^3\*z^3-2\*(2-\beta^2)\*\beta^4\*z^4-2\*\beta^5\*z^5\Big)\nn\\
&+&s\*(1-\beta^2)\*\Big(3+3\*\beta^4-4\*\beta^6-(1-8\*\beta^2+5\*\beta^4)\*\beta\*z-6\*(1-\beta^4)\*\beta^2\*z^2 \nn\\
&-&6\*(1-\beta^2)\*\beta^3\*z^3-2\*\beta^6\*z^4-2\*\beta^5\*z^5\Big)\Big]\*\Bomtmbmw\nn\\
&+&2\*\mb^2\*(1-\beta^2)\*\Cotmbmbmw\nn\\
&+&{s\*(1-\beta^2)+4\*(\mb^2-\mw^2)\over2\*(1-\beta\*z)}\* \nn\\
&&\Big(1+\beta\*(1-\beta^2)\*(2\*\beta-z-3\*\beta\*z^2)-\beta^4\*z^4\Big)\*\DBomtmbmw\Bigg\}\nn\\
&+&(z\ra-z),\end{eqnarray}
\begin{eqnarray}
{d\sigma^{V}_\chi\over dz} &=& 2\*{\alpha\over\pi}\*\sigma_0\*\chifac\*\genfacvert\*\Bigg\{{s\*(1-\beta^2)^2\over8}\nn\\
&+&{1\over2\*(1-\beta\*z)\*(1+2\*\beta\*z+\beta^2)}\* \nn\\
&&\Big(1+2\*\beta^2-\beta^4+(1+4\*\beta^2-3\*\beta^4)\*\beta\*z-2\*(1-\beta^2)\*(2\*\beta^2\*z^2+3\*\beta^3\*z^3)\nn\\
&-&2\*(1+\beta\*z)\*\beta^4\*z^4\Big)\*\Big(\Amt-\Amz\Big)\nn\\
&+&{(1-\beta^2)\over8\*(1-\beta^2\*z^2)\*(1+2\*\beta\*z+\beta^2)}\*\Big[2\*\mz^2\*\Big(1-4\*\beta^2-3\*\beta^4+4\*\beta^6\nn\\
&+&(1-8\*\beta^2+5\*\beta^4)\*\beta\*z+2\*(1+2\*\beta^2-3\*\beta^4)\*\beta^2\*z^2\nn\\
&+&6\*(1-\beta^2)\*\beta^3\*z^3+2\*\beta^6\*z^4+2\*\beta^5\*z^5\Big)\nn\\
&+&s\*\beta\*\Big(-(1+\beta^4)\*\beta+(1-7\*\beta^2+\beta^4+\beta^6)\*z+(3-12\*\beta^2+7\*\beta^4)\*\beta\*z^2\nn\\
&+&6\*(1-\beta^2)\*\beta^2\*z^3+2\*(4-3\*\beta^2)\*\beta^3\*z^4+4\*\beta^4\*z^5+2\*\beta^5\*z^6\Big)\Big]\*\Botmtmz\nn\\
&+&{1\over8\*(1-\beta^2\*z^2)}\*\Big[2\*\mz^2\*\Big(1+8\*\beta^2-11\*\beta^4+4\*\beta^6+(1+4\*\beta^2-3\*\beta^4)\*\beta\*z\nn\\
&-&2\*(5-8\*\beta^2+3\*\beta^4)\*\beta^2\*z^2-6\*(1-\beta^2)\*\beta^3\*z^3-2\*(2-\beta^2)\*\beta^4\*z^4-2\*\beta^5\*z^5\Big)\nn\\
&-&s\*(1-\beta^2)^2\*(1+2\*\beta\*z+\beta^2)\*(1-\beta\*z)\Big]\*\Bomtmtmz\nn\\
&-&{s^2\*(1-\beta^2)^2\*(1+2\*\beta\*z+\beta^2)\over16}\*\Cotmtmtmz\nn\\
&-&{\mz^2\*s\*(1-\beta^2)\over4\*(1-\beta\*z)}\nn\\
&&\*\Big(1+\beta\*(1-\beta^2)\*(2\*\beta-z-3\*\beta\*z^2)-\beta^4\*z^4\Big)\*\DBomtmtmz\Bigg\}\nn\\
&+&(z\ra-z),\end{eqnarray}
\begin{eqnarray}
{d\sigma^{V}_\phi\over dz} &=& {\alpha\over\pi}\*\sigma_0\*\genfacvert\*{\gw^2\over\mw^2}\*\Bigg\{{(1-\beta^2)\*(s\*(1-\beta^2)+4\*\mb^2)\over8}\nn\\
&+&{s\*(1-\beta^2)+4\*\mb^2\over2\*s\*(1-\beta\*z)\*(1+2\*\beta\*z+\beta^2)\*(1-\beta^2)}\*\Big(1+2\*\beta^2-\beta^4+(1+4\*\beta^2-3\*\beta^4)\*\beta\*z
\nn\\
&-&2\*(1-\beta^2)\*(2\*\beta^2\*z^2+3\*\beta^3\*z^3)-2\*(1+\beta\*z)\*\beta^4\*z^4\Big)\*\Big(\Amb-\Amw\Big)\nn\\
&+&{1\over16\*s\*(1-\beta^2\*z^2)\*(1+2\*\beta\*z+\beta^2)}\*\Big[\Big(4\*\mw^2\*(s\*(1-\beta^2)+4\*\mb^2)-16\*\mb^4\Big)\*\nn\\
&&\Big(1-4\*\beta^2-3\*\beta^4+4\*\beta^6+(1-8\*\beta^2+5\*\beta^4)\*\beta\*z+2\*(1+2\*\beta^2-3\*\beta^4)\*\beta^2\*z^2\nn\\
&+&6\*(1-\beta^2)\*\beta^3\*z^3+2\*\beta^6\*z^4+2\*\beta^5\*z^5\Big)\nn\\
&+&8\*\beta\*\mb^2\*s\*\Big(-(3-4\*\beta^2-\beta^4+4\*\beta^6)\*\beta + (1-11\*\beta^2+13\*\beta^4-7\*\beta^6)\*z\nn\\
&+&(5-18\*\beta^2+5\*\beta^4+6\*\beta^6)\*\beta\*z^2+2\*(5-11\*\beta^2+6\*\beta^4)\*\beta^2\*z^3 \nn\\
&+&2\*(5-3\*\beta^2-\beta^4)\*\beta^3\*z^4+4\*(2-\beta^2)\*\beta^4\*z^5+2\*\beta^5\*z^6\Big)\nn\\
&+&s^2\*(1-\beta^2)\*(1+2\*\beta\*z+\beta^2)\*\Big(1-4\*\beta^2-3\*\beta^4+4\*\beta^6+(1-8\*\beta^2+5\*\beta^4)\*\beta\*z\nn\\
&+&2\*(1+2\*\beta^2-3\*\beta^4)\*\beta^2\*z^2+6\*(1-\beta^2)\*\beta^3\*z^3+2\*\beta^6\*z^4+2\*\beta^5\*z^5\Big)\Big]\*\Botmbmw\nn\\
&+&{1\over16\*s\*(1-\beta^2)\*(1-\beta^2\*z^2)}\*\Big[\Big(4\*\mw^2\*(s\*(1-\beta^2)+4\*\mb^2)-16\*\mb^4\Big)\*\nn\\
&&\Big(1+8\*\beta^2-11\*\beta^4+4\*\beta^6+(1+4\*\beta^2-3\*\beta^4)\*\beta\*z-2\*(5-8\*\beta^2+3\*\beta^4)\*\beta^2\*z^2\nn\\
&-&6\*(1-\beta^2)\*\beta^3\*z^3-2\*(2-\beta^2)\*\beta^4\*z^4-2\*\beta^5\*z^5\Big)\nn\\
&-&8\*\mb^2\*s\*(1-\beta^2)^2\*\Big(1-\beta^2+4\*\beta^4+(1-\beta^2)\*\beta\*z-6\*\beta^4\*z^2+2\*\beta^4\*z^4\Big)\nn\\
&-&s^2\*(1-\beta^2)^2\*\Big(1-4\*\beta^2-3\*\beta^4+4\*\beta^6+(1-8\*\beta^2+5\*\beta^4)\*\beta\*z\nn\\
&+&2\*(1+2\*\beta^2-3\*\beta^4)\*\beta^2\*z^2+6\*(1-\beta^2)\*\beta^3\*z^3+2\*\beta^6\*z^4+2\*\beta^5\*z^5\Big)\Big]\*\Bomtmbmw\nn\\
&-&{1\over4}\*\mb^2\*(1-\beta^2)\*(s\*\beta^2+4\*s\*\beta\*z+3\*s-4\*\mb^2)\*\Cotmbmbmw\nn\\
&+&{1\over16\*(1-\beta\*z)}\*\Big(s^2\*(1-\beta^2)^2-4\*s\*(\mw^2+2\*\mb^2)\*(1-\beta^2)+16\*\mb^2\*(\mb^2-\mw^2)\Big)\nn\\
&&\*\Big(1+\beta\*(1-\beta^2)\*(2\*\beta-z-3\*\beta\*z^2)-\beta^4\*z^4\Big)\*\DBomtmbmw\Bigg\}\nn\\
&+&(z\ra-z),
\end{eqnarray}
\begin{eqnarray}
{d\sigma^{V}_H\over dz} &=& {\alpha\over\pi}\*\sigma_0\*{\gw^2\over\mw^2}\*\genfacvert\*\Bigg\{{s\*(1-\beta^2)^2\over8}\nn\\
&+&
{1\over2\*(1-\beta\*z)\*(1+2\*\beta\*z+\beta^2)}\*\Big(1+2\*\beta^2-\beta^4+(1+4\*\beta^2-3\*\beta^4)\*\beta\*z \nn\\
&-&2\*(1-\beta^2)\*(2\*\beta^2\*z^2+3\*\beta^3\*z^3)-2\*(1+\beta\*z)\*\beta^4\*z^4\Big)\*\Big(\Amt-\Amh\Big)\nn\\
&+&{1-\beta^2\over8\*(1-\beta^2\*z^2)\*(1+2\*\beta\*z+\beta^2)}\*\Big[2\*\mh^2\*\Big(1-4\*\beta^2-3\*\beta^4+4\*\beta^6\nn\\
&+&(1-8\*\beta^2+5\*\beta^4)\*\beta\*z+2\*(1+2\*\beta^2-3\*\beta^4)\*\beta^2\*z^2 \nn\\
&+&6\*(1-\beta^2)\*\beta^3\*z^3+2\*\beta^6\*z^4+2\*\beta^5\*z^5\Big)\nn\\
&+&s\*\beta\*\Big((3-8\*\beta^2-5\*\beta^4+8\*\beta^6)\*\beta+(1+\beta^2-23\*\beta^4+17\*\beta^6)\*z \nn\\
&-&(1-11\*\beta^4+12\*\beta^6)\*\beta\*z^2-2\*(1-13\*\beta^2+12\*\beta^4)\*\beta^2\*z^3\nn\\
&+&2\*(2-3\*\beta^2+2\*\beta^4)\*\beta^3\*z^4-4\*(1-2\*\beta^2)\*\beta^4\*z^5+2\*\beta^5\*z^6\Big)\Big]\*\Botmtmh\nn\\
&+&{1\over8\*(1-\beta^2\*z^2)}\*\Big[2\*\mh^2\*\Big(1+8\*\beta^2-11\*\beta^4+4\*\beta^6+(1+4\*\beta^2-3\*\beta^4)\*\beta\*z\nn\\
&-&2\*(5-8\*\beta^2+3\*\beta^4)\*\beta^2\*z^2-6\*(1-\beta^2)\*\beta^3\*z^3-2\*(2-\beta^2)\*\beta^4\*z^4-2\*\beta^5\*z^5\Big)\nn\\
&-&s\*(1-\beta^2)^2\*\Big(1+5\*\beta^2-8\*\beta^4+(1-\beta^2)\*\beta\*z-6\*(1-2\*\beta^2)\*\beta^2\*z^2-4\*\beta^4\*z^4\Big)\Big]\*\Bomtmtmh\nn\\
&+&{s^2\*(1-\beta^2)^2\*(3+2\*\beta\*z-\beta^2)\over16}\*\Cotmtmtmh\nn\\
&+&{s\*(1-\beta^2)\*(s\*(1-\beta^2)-\mh^2)\over4\*(1-\beta\*z)}\*\Big(1+\beta\*(1-\beta^2)\*(2\*\beta-z-3\*\beta\*z^2)-\beta^4\*z^4)\nn\\
&&\*\DBomtmtmh\Bigg\}+(z\ra-z).
\end{eqnarray} 
Box corrections:
\begin{eqnarray}
{d\sigma^{\Box}_Z\over dz} &=&  {\alpha\over8\*\pi}\*\sigma_0\*\genfacbox\*\Bigg\{-2\*(\gvt^2+\gat^2)\*\beta\*z\*(1-z^2) \nn\\
&+&8\*(\gvt^2+\gat^2)\*{\beta\*(1-z^2)\*(4\*\beta-z-2\*\beta\*z^2)\over s\*(1+\beta^2+2\*\beta\*z)}\*\Big(\Amt-\Amz\Big) \nn\\
&-&2\*\Big[2\*(\gvt^2+\gat^2)\*{\mz^2\over s\*\beta}\*z\*(5-4\*\beta^2-(3-2\*\beta^2)\*z^2) \nn\\
&-&\gvt^2\*(2+(1-4\*\beta^2)\*\beta\*z-2\*z^2+(1+2\*\beta^2)\*\beta\*z^3)\nn\\
&-&\gat^2\*(2-3\*(5-4\*\beta^2)\*\beta\*z-2\*z^2+3\*(3-2\*\beta^2)\*\beta\*z^3)\Big]\*
\Bosmtmt \nn\\
&+&{4\over 1+\beta\*z}\*\Big[-(\gat^2+\gvt^2)\*{\mz^2\over s\*\beta}\*\Big(-6\*\beta^3-(5-8\*\beta^2+4\*\beta^4)\*z \nn\\
&-&(5-12\*\beta^2)\*\beta\*z^2+(3-4\*\beta^2+2\*\beta^4)\*z^3+(3-4\*\beta^2)\*\beta\*z^4\Big)\nn\\
&+&\gvt^2\*\Big(5\*\beta^2-3\*\beta^4+(1+\beta^2-2\*\beta^4)\*\beta\*z+(1-4\*\beta^2+3\*\beta^4)\*z^2\nn \\
&-&(1-\beta^2)\*\beta^3\*z^3-(1+\beta^2)\*\beta^2\*z^4\Big)\nn\\
&-&\gat^2\*\Big(3\*\beta^2-5\*\beta^4-3\*(3-5\*\beta^2+2\*\beta^4)\*\beta\*z-(1+8\*\beta^2-9\*\beta^4)\*z^2\nn\\
&+&(4-7\*\beta^2+3\*\beta^4)\*\beta\*z^3+(5-3\*\beta^2)\*\beta^2\*z^4\Big)\Big]\*\Bomtmtmz \nn\\
&+&2\*\Big[8\*(\gvt^2+\gat^2)\*{\mz^4\over s}+4\*\mz^2\*\Big(\gvt^2\*(3+\beta\*z)-\gat^2\*(1-4\*\beta^2-\beta\*z)\Big) \nn\\
&+&s\*\Big(\gvt^2\*(4+2\*\beta^2-\beta^4+2\*(2-\beta^2)\*\beta\*z+\beta^2\*z^2)\nn\\
&+&\gat^2\*(4-6\*\beta^2+7\*\beta^4-2\*(2-3\*\beta^2)\*\beta\*z+\beta^2\*z^2)\Big)\Big]\*\Cosmtmtmt\nn\\
&+&2\*\Big[-2\*(\gvt^2+\gat^2)\*{\mz^4\over s\*\beta}\*z\*(5-8\*\beta^2-3\*z^2+2\*\beta^2\*z^2) \nn\\
&+&2\*\mz^2\*\Big(\gvt^2\*(1+\beta^2+4\*\beta\*z-(1-\beta^2)\*z^2+2\*\beta\*z^3)\nn\\
&+&\gat^2\*(1+\beta^2-4\*(3-4\*\beta^2)\*\beta\*z-(1-\beta^2)\*z^2+2\*(3-2\*\beta^2)\*\beta\*z^3)\Big)\nn\\
&+&\beta\*s\*\Big(\gvt^2\*((4-\beta^2)\*\beta+\beta^4\*z^3+4\*(1+\beta^2-\beta^4)\*z-\beta^3\*z^2+\beta^2\*z^3)\nn\\
&-&\gat^2\*(-3\*\beta^3-4\*(1-3\*\beta^2+3\*\beta^4)\*z+(4-3\*\beta^2)\*\beta\*z^2-(5-3\*\beta^2)\*\beta^2\*z^3)\Big)\Big]\*\Cosmtmtmz \nn\\
&+&{4\over (1+\beta\*z)\*(1+\beta^2+2\*\beta\*z)}\nn\\
&&\*\Big[2\*(\gvt^2+\gat^2)\*{\mz^2\over s}\*\beta\*(z+\beta)\*\Big(1-3\*\beta^2+(1-2\*\beta^2)\*\beta\*z
+2\*\beta^2\*z^2+\beta^3\*z^3\Big)\nn\\
&-&\gvt^2\*\Big(1+8\*\beta^2-3\*\beta^6+(4+14\*\beta^2-11\*\beta^4-2\*\beta^6)\*\beta\*z \nn\\
&-&(2+\beta^2+3\*\beta^4)\*\beta^2\*z^2-(11-6\*\beta^2-\beta^4)\*\beta^3\*z^3-2\*(1-\beta^2)\*\beta^4\*z^4-\beta^5\*z^5\Big) \nn\\
&-&\gat^2\*\Big(1+5\*\beta^6+(4-10\*\beta^2+5\*\beta^4+6\*\beta^6)\*\beta\*z+(2-17\*\beta^2+9\*\beta^4)\*\beta^2\*z^2\nn\\
&+&(1-2\*\beta^2-3\*\beta^4)\*\beta^3\*z^3+6\*(1-\beta^2)\*\beta^4\*z^4-\beta^5\*z^5\Big)\Big]\*\Botmtmz\nn\\
&-&2\*\Big[8\*(\gvt^2+\gat^2)\*{\mz^4\over s}\*(1+\beta\*z)\nn\\
&+& 4\*\mz^2\*(1+\beta\*z)\*\Big(3\*\gvt^2-\gat^2+4\*\beta^2\*\gat^2+(\gvt^2+\gat^2)\*\beta\*z\Big) \nn\\
&+&s\*\Big(\gvt^2\*(4+2\*\beta^2-\beta^4+(7+2\*\beta^2-2\*\beta^4)\*\beta\*z+(5-2\*\beta^2)\*\beta^2\*z^2+\beta^3\*z^3) \nn\\
&+&\gat^2\*(4-6\*\beta^2+7\*\beta^4-(1-2\*\beta^2-6\*\beta^4)\*\beta\*z-3\*(1-2\*\beta^2)\*\beta^2\*z^2+\beta^3\*z^3)\Big)\Big]\*\Cotmtmtmz\nn\\
&+&\Big[16\*(\gvt^2+\gat^2)\*{\mz^6\over s}+8\*\mz^4\*(\gvt^2\*(3+\beta^2+2\*\beta\*z)-\gat^2\*(1-5\*\beta^2-2\*\beta\*z)) \nn\\
&+&2\*s\*\mz^2\*\Big(\gvt^2\*(4+9\*\beta^2-2\*\beta^4+10\*\beta\*z+(2+\beta^2)\*\beta^2\*z^2)\nn\\
&+&\gat^2\*(4-7\*\beta^2+14\*\beta^4-2\*(3-8\*\beta^2)\*\beta\*z+(2+\beta^2)\*\beta^2\*z^2)\Big) \nn\\ 
&+&s^2\*\Big(\gvt^2\*(1+7\*\beta^2-\beta^4-2\*\beta^6+(3+8\*\beta^2-4\*\beta^4)\*\beta\*z\nn\\
&+&(1+3\*\beta^2-\beta^4)\*\beta^2\*z^2+\beta^3\*z^3)-\gat^2\*(3-7\*\beta^2+5\*\beta^4-6\*\beta^6 \nn\\
&-&(3-8\*\beta^2+12\*\beta^4)\*\beta\*z+3\*(1-\beta^2-\beta^4)\*\beta^2\*z^2-\beta^3\*z^3)\Big)\Big]\*\Dotz\Bigg\}\nn\\
&+&(z\ra-z),\end{eqnarray}
\begin{eqnarray}
{d\sigma^{\Box}_W\over dz} &=& 
{\alpha\over2\*\pi}\*\sigma_0\*\gw^2\*\genfacbox\*\Bigg\{-(1-z^2)\*\beta\*z\nn\\
&+&{4\*\beta\*(1-z^2)\*(4\*\beta-z-2\*\beta\*z^2)\over s\*(1+\beta^2+2\*\beta\*z)}\*\Big(\Amb-\Amw\Big)\nn\\
&-&{1\over2\*s\*\beta}\*\Big[4\*(\mw^2-\mb^2)\*z\*(5-4\*\beta^2-(3-2\*\beta^2)\*z^2)\nn\\
&-&s\*\Big(4\*\beta-(5+5\*\beta^2-4\*\beta^4)\*z-4\*\beta\*z^2+(3+5\*\beta^2-2\*\beta^4)\*z^3\Big)\Big]\*\Bosmbmb \nn\\
&+&{1\over2\*s\*\beta\*(1+\beta\*z)}\*\Big[4\*(\mw^2-\mb^2)\*\nn\\
&&\Big(6\*\beta^3+(5-8\*\beta^2+4\*\beta^4)\*z+(5-12\*\beta^2)\*\beta\*z^2-(3-4\*\beta^2+2\*\beta^4)\*z^3\nn\\
&-&(3-4\*\beta^2)\*\beta\*z^4\Big)+s\*\Big(2\*(5-\beta^2)\*\beta^3+(1-\beta^2)\*((5+12\*\beta^2-4\*\beta^4)\*z\nn\\
&+&9\*\beta\*z^2-(3+4\*\beta^2-2\*\beta^4)\*z^3)-\beta\*(3+5\*\beta^2)\*z^4\Big)\Big]\*\Bomtmbmw\nn\\
&+&{1\over2\*s}\*\Big[(4\*(\mw^2-\mb^2))^2+8\*\mw^2\*s\*(2+\beta^2+\beta\*z)-8\*\mb^2\*s\*\beta\*(2\*\beta+z)\nn\\
&+& s^2\*(7+2\*\beta^2+\beta^4+2\*(1+\beta^2)\*\beta\*z +2\*\beta^2\*z^2)\Big]\*\Cosmbmbmb\nn\\
&+&{1\over8\*s\*\beta}\*\Big[-(4\*(\mw^2-\mb^2))^2\*z\*(5-8\*\beta^2-(3-2\*\beta^2)\*z^2) \nn\\
&+&8\*\mw^2\*s\*\Big(2\*(1+\beta^2)\*\beta-(5-5\*\beta^2-8\*\beta^4)\*z\nn\\
&-&2\*(1-\beta^2)\*\beta\*z^2+(3+3\*\beta^2-2\*\beta^4)\*z^3\Big) \nn\\
&-&8\*\mb^2\*s\*\Big(2\*(1+\beta^2)\*\beta-(1+\beta^2)\*(5-8\*\beta^2)\*z\nn\\
&-&2\*(1-\beta^2)\*\beta\*z^2+(1+\beta^2)\*(3-2\*\beta^2)\*z^3\Big) \nn\\
&-&s^2\*\Big(-4\*(1+4\*\beta^2+\beta^4)\*\beta+(5-30\*\beta^2+\beta^4-8\*\beta^6)\*z\nn\\
&+&4\*(1+2\*\beta^2-\beta^4)\*\beta\*z^2-(3+4\*\beta^2+11\*\beta^4-2\*\beta^6)\*z^3\Big)\Big]\*\Cosmbmbmw\nn\\
&+&{1\over s\*(1+\beta\*z)\*(1+\beta^2+2\*\beta\*z)}\*\Big[\nn\\
&&4\*(\mw^2-\mb^2)\*\beta\*(z+\beta)\*
(1-3\*\beta^2+(1-2\*\beta^2)\*\beta\*z+2\*\beta^2\*z^2+\beta^3\*z^3)\nn\\
&-&s\*(1+\beta^2+2\*\beta\*z)\*\Big(2+5\*\beta^2-\beta^4+(3-6\*\beta^2+2\*\beta^4)\*\beta\*z\nn\\
&-&(7-2\*\beta^2)\*\beta^2\*z^2+(2-\beta^2)\*\beta^3\*z^3-\beta^4\*z^4\Big)\Big]\*\Botmbmw \nn\\
&-&{1\over2\*s}\*\Big[(4\*(\mw^2-\mb^2))^2\*(1+\beta\*z)+8\*\mw^2\*s\*(1+\beta\*z)\*(2+\beta^2+\beta\*z)\nn\\
&-&8\*\mb^2\*s\*\beta\*(\beta+z)\*(2+\beta\*z)\nn\\
&+&s^2\*(1+\beta\*z)\*\Big(7+2\*\beta^2+\beta^4+2\*(1+\beta^2)\*\beta\*z+2\*\beta^2\*z^2\Big)\Big]\*\Cotmbmbmw\nn\\
&+&{1\over8\*s}\*\Big[(4\*(\mw^2-\mb^2))^3+8\*s\*\Big(2\*(\mw^2-\mb^2)^2\*(3\*\beta^2+4\*\beta\*z)\nn\\
&+&10\*\mw^4+2\*\mb^4\*(1+2\*\beta^2\*z^2)-4\*\mw^2\*\mb^2\*(3+\beta^2\*z^2)\Big)\nn\\
&+&4\*s^2\*\Big(\mw^2\*(11+8\*\beta^2+3\*\beta^4+4\*(3+2\*\beta^2)\*\beta\*z+6\*\beta^2\*z^2)\nn\\
&-&\mb^2\*(11-4\*\beta^2+3\*\beta^4+8\*(1+\beta^2)\*\beta\*z+2\*(6+\beta^2)\*\beta^2\*z^2)\Big)  \nn\\
&+&s^3\*(1+\beta^2+2\*\beta\*z)\*(7+2\*\beta^2+\beta^4+2\*(1+\beta^2)\*\beta\*z+2\*\beta^2\*z^2)\Big]\*\Dotw\Bigg\}\nn\\
&+&(z\ra-z),\end{eqnarray}
\begin{eqnarray}
{d\sigma^{\Box}_{\chi}\over dz}  &=&
{\alpha\over2\*\pi}\*\sigma_0\*\mt^2\*\chifac\*\genfacbox\*\Bigg\{-\beta\*z\*(1-z^2)\nn\\
&+&{4\*\beta\*(1-z^2)\*(4\*\beta-z-2\*\beta\*z^2)\over s\*(1+\beta^2+2\*\beta\*z)}\*\Big(\Amt-\Amz\Big)\nn\\
&-&{1\over s\*\beta}\*\Big[2\*\mz^2\*z\*(5-4\*\beta^2-(3-2\*\beta^2)\*z^2)-s\*\beta\*(1-z^2)\*(2+\beta\*z)\Big]\*\Bosmtmt\nn\\
&+&{1\over s\*\beta\*(1+\beta\*z)}\*\Big[2\*\mz^2\*(6\*\beta^3+(5-8\*\beta^2+4\*\beta^4)\*z+(5-12\*\beta^2)\*\beta\*z^2\nn\\
&-&(3-4\*\beta^2+2\*\beta^4)\*z^3-(3-4\*\beta^2)\*\beta\*z^4\Big)\nn\\
&-&2\*\beta\*s\*(1-\beta^2)\*(2-\beta^2+\beta\*z-z^2-\beta\*z^3)\Big]\*\Bomtmtmz\nn\\
&+&{1\over s}\*\Big[8\*\mz^4+4\*\beta\*s\*(\beta+z)\*\mz^2+s^2\*(2-2\*\beta^2+\beta^4+2\*\beta\*z+\beta^2\*z^2)\Big]\*\Cosmtmtmt\nn\\
&-&{1\over s\*\beta}\*\Big[2\*\mz^4\*z\*(5-8\*\beta^2-(3-2\*\beta^2)\*z^2)\nn\\
&-&2\*\mz^2\*s\*\beta\*(1+\beta^2-2\*(1-2\*\beta^2)\*\beta\*z-(1-\beta^2)\*z^2+(1-\beta^2)\*\beta\*z^3)\nn\\
&-&s^2\*\beta^2\*(\beta+z)\*(1+\beta\*z)\Big]\*\Cosmtmtmz\nn\\
&+&{1\over s\*(1+\beta\*z)\*(1+\beta^2+2\*\beta\*z)}\*\Big[
4\*\mz^2\*\beta\*(\beta+z)\*\Big(1-3\*\beta^2+(1-2\*\beta^2)\*\beta\*z+2\*z^2\*\beta^2+\beta^3\*z^3\Big)
\nn\\
&+&2\*s\*\Big(1-4\*\beta^2+\beta^6+(2-10\*\beta^2+3\*\beta^4)\*\beta\*z+(3-5\*\beta^2)\*\beta^2\*z^2\nn\\
&+&2\*(3-\beta^2)\*\beta^3\*z^3+4\*\beta^4\*z^4+\beta^5\*z^5\Big)\Big]\*\Botmtmz\nn\\
&-&{1\over s}\*
\Big[4\*\Big(2\*\mz^4+\mz^2\*s\*\beta\*(\beta+z)\Big)\*(1+\beta\*z)\nn\\
&+&s^2\*\Big(2-2\*\beta^2+\beta^4+3\*\beta\*z+3\*\beta^2\*z^2+\beta^3\*z^3\Big)\Big]\*\Cotmtmtmz\nn\\
&+&{1\over2\*s}\*\Big[16\*\mz^6+16\*\mz^4\*s\*\beta\*(\beta+z)\nn\\
&+&2\*\mz^2\*s^2\*\Big(2-\beta^2+2\*\beta^4+2\*(1+2\*\beta^2)\*\beta\*z+(2+\beta^2)\*\beta^2\*z^2\Big)\nn\\
&+&s^3\*\beta\*(\beta+z)\*(1+\beta\*z)^2\Big]\*\Dotz\Bigg\}+(z\ra-z),\end{eqnarray}
\begin{eqnarray}
{d\sigma^{\Box}_{\phi}\over dz} &=& {\alpha\over4\*\pi}\*\sigma_0\*{\gw^2\over \mw^2}\*\genfacbox\*\Bigg\{-{\beta\over4}\*z\*(1-z^2)\*(s\*(1-\beta^2)+4\*\mb^2)\nn\\
&+&{\beta\*(1-z^2)\*(s\*(1-\beta^2)+4\*\mb^2)\*(4\*\beta-z-2\*\beta\*z^2)\over s\*(1+\beta^2+2\*\beta\*z)}\*\Big(\Amb-\Amw\Big)\nn\\
&+&{1\over8\*s\*\beta}\*\Big[(16\*\mb^2\*(\mb^2-\mw^2)-4\*\mw^2\*s\*(1-\beta^2))\*z\*(5-4\*\beta^2-(3-2\*\beta^2)\*z^2)\nn\\
&+&8\*\mb^2\*s\*\beta\*(2+(5-4\*\beta^2)\*\beta\*z-2\*z^2-(3-2\*\beta^2)\*\beta\*z^3)\nn\\
&+&s^2\*(1-\beta^2)\*(4\*\beta-(5-3\*\beta^2-4\*\beta^4)\*z-4\*\beta\*z^2+(3-3\*\beta^2-2\*\beta^4)\*z^3)\Big]\*\Bosmbmb\nn\\
&-&{1\over8\*s\*\beta\*(1+\beta\*z)}\*\Big[\Big(16\*\mb^2\*(\mb^2-\mw^2)-4\*\mw^2\*s\*(1-\beta^2)\Big)\*\nn\\
&&\Big(6\*\beta^3+(5-8\*\beta^2+4\*\beta^4)\*z+(5-12\*\beta^2)\*\beta\*z^2\nn\\
&-&(3-4\*\beta^2+2\*\beta^4)\*z^3-(3-4\*\beta^2)\*\beta\*z^4\Big)\nn\\
&+&16\*\mb^2\*s\*\beta\*(1-\beta^2)\*\Big(2-3\*\beta^2+(3-2\*\beta^2)\*\beta\*z-(1-3\*\beta^2)\*z^2\nn\\
&-&(2-\beta^2)\*\beta\*z^3-\beta^2\*z^4\Big)+s^2\*(1-\beta^2)^2\*\Big(8\*\beta-2\*\beta^3-(5-4\*\beta^2-4\*\beta^4)\*z\nn\\
&-&9\*\beta\*z^2+(3-4\*\beta^2-2\*\beta^4)\*z^3+3\*\beta\*z^4\Big)\Big]\*\Bomtmbmw\nn\\
&+&{1\over8\*s}\*\Big[64\*\mb^6+16\*\mb^4\*(s\*(1-\beta^2)-8\*\mw^2-2\*z\*s\*\beta)\nn\\
&+&16\*\mw^4\*(s\*(1-\beta^2)+4\*\mb^2)+32\*\mb^2\*\mw^2\*s\*(1+\beta\*z)\nn\\
&+&4\*s^2\*\mb^2\*\Big(3-2\*\beta^2+\beta^4+4\*(2-\beta^2)\*\beta\*z+2\*\beta^2\*z^2\Big)+8\*s^2\*\mw^2\*\beta\*(1-\beta^2)\*(\beta+z)\nn\\
&+&s^3\*(1-\beta^2)\*(3-2\*\beta^2+\beta^4+2\*(1+\beta^2)\*\beta\*z+2\*\beta^2\*z^2)\Big]\*\Cosmbmbmb\nn\\
&+&{1\over32\*s\*\beta}\*\Big[-16\*\Big(4\*\mb^6+\mw^4\*(s\*(1-\beta^2)+4\*\mb^2)-8\*\mb^4\*\mw^2\Big)\nn\\
&&\*z\*\Big(5-8\*\beta^2-(3-2\*\beta^2)\*z^2\Big)\nn\\
&-&16\*\mb^4\*s\*\Big(4\*(1+\beta^2)\*\beta-(5-17\*\beta^2+8\*\beta^4)\*z\nn\\
&-&4\*(1-\beta^2)\*\beta\*z^2+(3-9\*\beta^2+2\*\beta^4)\*z^3\Big)\nn\\
&+&64\*\mw^2\*\mb^2\*s\*\beta\*\Big(1+\beta^2+2\*\beta\*z-(1-\beta^2)\*z^2\Big)\nn\\
&+&4\*s^2\*\mb^2\*\Big(8\*(2-\beta^2)\*\beta^3+(5-2\*\beta^2+21\*\beta^4-8\*\beta^6)\*z\nn\\
&+&8\*(2-\beta^2)\*\beta^3\*z^2-(3-\beta^4-2\*\beta^6)\*z^3\Big) \nn\\
&+&8\*\mw^2\*s^2\*(1-\beta^2)\*\Big(2\*(1+\beta^2)\*\beta-(5-\beta^2-8\*\beta^4)\*z\nn\\
&-&2\*(1-\beta^2)\*\beta\*z^2+(3-\beta^2-2\*\beta^4)\*z^3\Big)\nn\\
&+&s^3\*(1-\beta^2)\*\Big(4\*(1+\beta^4)\*\beta-(5-22\*\beta^2+9\*\beta^4-8\*\beta^6)\*z\nn\\
&-&4\*(1-2\*\beta^2-\beta^4)\*\beta\*z^2+(3-4\*\beta^2+3\*\beta^4-2\*\beta^6)\*z^3\Big)\Big]\*\Cosmbmbmw\nn\\
&-&{1\over4\*s\*(1+\beta\*z)\*(1+\beta^2+2\*\beta\*z)}\*\Big[\Big(16\*\mb^2\*(\mb^2-\mw^2)-4\*\mw^2\*s\*(1-\beta^2)\Big)\nn\\
&&\*\beta\*(\beta+z)\*\Big(1-3\*\beta^2+(1-2\*\beta^2)\*\beta\*z+2\*\beta^2\*z^2+\beta^3\*z^3\Big)\nn\\
&-&8\*\mb^2\*s\*\Big(1-6\*\beta^2+3\*\beta^6+(2-16\*\beta^2+7\*\beta^4+2\*\beta^6)\*\beta\*z \nn\\
&+&(4-9\*\beta^2+3\*\beta^4)\*\beta^2\*z^2+(9-4\*\beta^2-\beta^4)\*\beta^3\*z^3+2\*(3-\beta^2)\*\beta^4\*z^4+\beta^5\*z^5\Big)\nn\\
&-&s^2\*(1+\beta^2+2\*\beta\*z)\*(1-\beta^2)\*\Big(2-5\*\beta^2+\beta^4+(1-2\*\beta^2-2\*\beta^4)\*\beta\*z\nn\\
&+&(3-2\*\beta^2)\*\beta^2\*z^2+(2+\beta^2)\*\beta^3\*z^3+\beta^4\*z^4\Big)\Big]\*\Botmbmw\nn\\
&-&{1\over8\*s}\*\Big[64\*\mb^6\*(1+\beta\*z)-16\*\mb^4\*\Big(8\*\mw^2\*(1+\beta\*z)-s\*(1-\beta^2-(3-\beta^2)\*\beta\*z-2\*\beta^2\*z^2)\Big)\nn\\
&+&\mb^2\*\Big(32\*(2\*\mw^4+\mw^2\*s\*(1+\beta\*z))\*(1+\beta\*z)\nn\\
&+&4\*s^2\*(3-4\*\beta^4\*z^2+9\*\beta\*z-\beta^5\*z+2\*\beta^3\*z^3+10\*\beta^2\*z^2-2\*\beta^3\*z-2\*\beta^2+\beta^4)\Big)\nn\\
&+&16\*\mw^4\*s\*(1-\beta^2)\*(1+\beta\*z)+8\*\mw^2\* s^2\*\beta\*(1-\beta^2)\*(\beta+z)\*(1+\beta\*z)\nn\\
&+&s^3\*(1-\beta^2)\*(1+\beta\*z)\*\Big(3-2\*\beta^2+\beta^4+2\*(1+\beta^2)\*\beta\*z+2\*\beta^2\*z^2\Big)\Big]\*\Cotmbmbmw\nn\\
&-&{1\over32\*s}\*\Big[256\*\mb^8-128\*\mb^6\*(6\*\mw^2+s\*\beta\*z\*(2+\beta\*z))\nn\\
&-&64\*\mw^6\*s\*(1-\beta^2)+32\*\mb^4\*\Big(24\*\mw^4+2\*\mw^2\*s\*(3+\beta^2+8\*\beta\*z+2\*\beta^2\*z^2)\nn\\
&-&s^2\*(-3\*\beta^2+\beta^4+1+2\*\beta^4\*z^2-5\*\beta^2\*z^2-6\*\beta\*z+2\*\beta^3\*z)\Big)\nn\\
&-&16\*\mw^4\*s^2\*(1-\beta^2)\*(1+3\*\beta^2+4\*\beta\*z)\nn\\
&-&8\*\mb^2\*\Big(32\*\mw^6+16\*\mw^4\*s\*(1+\beta^2+2\*\beta\*z)+2\*\mw^2\*s^2\*\big(5+\beta^4+12\*\beta\*z+2\*(2+\beta^2)\*\beta^2\*z^2\Big)\nn\\
&+&s^3\*\Big(4\*\beta^3\*z-4\*\beta^4\*z^2+\beta^6\*z^2-2\*\beta^5\*z+6\*\beta^2-2+4\*\beta\*z+2\*\beta^3\*z^3+9\*\beta^2\*z^2-2\*\beta^4\Big)\Big)\nn\\
&-&4\*\mw^2\*s^3\*(1-\beta^2)\*(3+3\*\beta^4+4\*(1+2\*\beta^2)\*\beta\*z+6\*\beta^2\*z^2)\nn\\
&-&s^4\*(1-\beta^2)\*(1+\beta^2+2\*\beta\*z)\*(3-2\*\beta^2+\beta^4+2\*(1+\beta^2)\*\beta\*z+2\*\beta^2\*z^2)\Big]\*\Dotw
\Bigg\}\nn\\
&+&(z\ra-z),\end{eqnarray}
\begin{eqnarray}
{d\sigma^{\Box}_{H}\over dz}  &=&
-{\alpha\over4\*\pi}\*\sigma_0\*\gw^2\*{\mt^2\over \mw^2}\*\genfacbox\*\Bigg\{\beta\*z\*(1-z^2)\nn\\
&-&{4\*\beta\*(1-z^2)\*(4\*\beta-z-2\*\beta\*z^2)\over s\*(1+\beta^2+2\*\beta\*z)}\*\Big(\Amt-\Amh\Big)\nn\\
&+&{1\over s\*\beta}\*\Big[2\*\mh^2\*z\*(5-4\*\beta^2-(3-2\*\beta^2)\*z^2)\nn\\
&-&s\*\beta\*\Big(2-(7-8\*\beta^2)\*\beta\*z-2\*z^2+(3-4\*\beta^2)\*\beta\*z^3\Big)\Big]\*\Bosmtmt\nn\\
&+&{1\over s\*\beta\*(1+\beta\*z)}\*\Big[2\*\mh^2\*\Big(-6\*\beta^3-(5-8\*\beta^2+4\*\beta^4)\*z\nn\\
&-&(5-12\*\beta^2)\*\beta\*z^2+(3-4\*\beta^2+2\*\beta^4)\*z^3+(3-4\*\beta^2)\*\beta\*z^4\Big)\nn\\
&+&2\*s\*\beta\*(1-\beta^2)\*\Big(2+3\*\beta^2-(3-4\*\beta^2)\*\beta\*z \nn\\
&-&(1+6\*\beta^2)\*z^2+(1-2\*\beta^2)\*\beta\*z^3+2\*\beta^2\*z^4\Big)\Big]\*\Bomtmtmh\nn\\
&-&{1\over s}\*\Big[8\*\mh^4-4\*\mh^2\*s\*(2-3\*\beta^2-\beta\*z)\nn\\
&+&s^2\*(6-10\*\beta^2+5\*\beta^4-2\*(1-2\*\beta^2)\*\beta\*z+\beta^2\*z^2)\Big]\*\Cosmtmtmt\nn\\
&+&{1\over s\*\beta}\*\Big[2\*\mh^4\*z\*(5-8\*\beta^2-(3-2\*\beta^2)\*z^2)\nn\\
&-&2\*\mh^2\*s\*\beta\*\Big(1+\beta^2-2\*(5-6\*\beta^2)\*\beta\*z-(1-\beta^2)\*z^2+3\*(1-\beta^2)\*\beta\*z^3\Big)\nn\\
&+&s^2\*\beta^2\*\Big(\beta-2\*\beta^3-(7-13\*\beta^2+8\*\beta^4)\*z \nn\\
&+&(1-2\*\beta^2)\*\beta\*z^2-2\*(1-\beta^2)\*\beta^2\*z^3\Big)\Big]\*\Cosmtmtmh\nn\\
&+&{1\over s\*(1+\beta\*z)\*(1+\beta^2+2\*\beta\*z)}\*\nn\\
&&\Big[-4\*\mh^2\*\beta\*(\beta+z)\*\Big(1-3\*\beta^2+(1-2\*\beta^2)\*\beta\*z+2\*\beta^2\*z^2+\beta^3\*z^3\Big)\nn\\
&-&2\*s\*\Big(1-3\*\beta^6+(2+2\*\beta^2-5\*\beta^4-4\*\beta^6)\*\beta\*z+(1+3\*\beta^2-6\*\beta^4)\*\beta^2\*z^2\nn\\
&+&2\*(1+\beta^2)\*\beta^5\*z^3+4\*\beta^6\*z^4+\beta^5\*z^5\Big)\Big]\*\Botmtmh\nn\\
&+&{1\over s}\*\Big[\Big(8\*\mh^4-4\*\mh^2\*s\*(2-3\*\beta^2-\beta\*z)\Big)\*(1+\beta\*z)\nn\\
&+&s^2\*(6-10\*\beta^2+5\*\beta^4+(3-4\*\beta^2+4\*\beta^4)\*\beta\*z-(1-4\*\beta^2)\*\beta^2\*z^2+\beta^3\*z^3)\Big]\*\Cotmtmtmh\nn\\
&+&{1\over 2\*s}\*\Big[-16\*\mh^6+16\*\mh^4\*s\*(1-2\*\beta^2-\beta\*z)\nn\\
&-&2\*\mh^2\*s^2\*\Big(6-13\*\beta^2+10\*\beta^4-6\*(1-2\*\beta^2)\*\beta\*z+(2+\beta^2)\*\beta^2\*z^2\Big)\nn\\
&-&s^3\*\Big(2+\beta^2-6\*\beta^4+4\*\beta^6+(5-10\*\beta^2+8\*\beta^4)\*\beta\*z\nn\\
&+&(1+2\*\beta^2)\*\beta^4\*z^2+\beta^3\*z^3\Big)\Big]\*\Doth
\Bigg\}+(z\ra-z).\end{eqnarray}
\pagebreak


\begin{thebibliography}{10}

\bibitem{Nason:1988xz}
P. Nason, S. Dawson and R.K. Ellis,
\newblock Nucl. Phys. B303 (1988) 607,
\newblock 

\bibitem{Nason:1989zy}
P. Nason, S. Dawson and R.K. Ellis,
\newblock Nucl. Phys. B327 (1989) 49,
\newblock 

\bibitem{Beenakker:1989bq}
W. Beenakker et~al.,
\newblock Phys. Rev. D40 (1989) 54,
\newblock 

\bibitem{Beenakker:1991ma}
W. Beenakker et~al.,
\newblock Nucl. Phys. B351 (1991) 507,
\newblock 

\bibitem{Bernreuther:2001rq}
W. Bernreuther et~al.,
\newblock Phys. Rev. Lett. 87 (2001) 242002,
  \href{http://www.arXiv.org/abs/hep-ph/0107086}{{\tt hep-ph/0107086}},
\newblock 

\bibitem{Laenen:1992af}
E. Laenen, J. Smith and W.L. van Neerven,
\newblock Nucl. Phys. B369 (1992) 543,
\newblock 

\bibitem{Kidonakis:1995wz}
N. Kidonakis and J. Smith,
\newblock Phys. Rev. D51 (1995) 6092,
  \href{http://www.arXiv.org/abs/hep-ph/9502341}{{\tt hep-ph/9502341}},
\newblock 

\bibitem{Berger:1996ad}
E.L. Berger and H. Contopanagos,
\newblock Phys. Rev. D54 (1996) 3085,
  \href{http://www.arXiv.org/abs/hep-ph/9603326}{{\tt hep-ph/9603326}},
\newblock 

\bibitem{Catani:1996yz}
S. Catani et~al.,
\newblock Nucl. Phys. B478 (1996) 273,
  \href{http://www.arXiv.org/abs/hep-ph/9604351}{{\tt hep-ph/9604351}},
\newblock 

\bibitem{Berger:1998gz}
E.L. Berger and H. Contopanagos,
\newblock Phys. Rev. D57 (1998) 253,
  \href{http://www.arXiv.org/abs/hep-ph/9706206}{{\tt hep-ph/9706206}},
\newblock 

\bibitem{Cacciari:2003fi}
M. Cacciari et~al.,
\newblock JHEP 04 (2004) 068,
  \href{http://www.arXiv.org/abs/hep-ph/0303085}{{\tt hep-ph/0303085}},
\newblock 

\bibitem{Bernreuther:2004jv}
W. Bernreuther et~al.,
\newblock Nucl. Phys. B690 (2004) 81,
  \href{http://www.arXiv.org/abs/hep-ph/0403035}{{\tt hep-ph/0403035}},
\newblock 

\bibitem{Kuhn:1999nn}
J.H. Kühn, A.A. Penin and V.A. Smirnov,
\newblock Eur. Phys. J. C17 (2000) 97,
  \href{http://www.arXiv.org/abs/hep-ph/9912503}{{\tt hep-ph/9912503}},
\newblock 

\bibitem{Kuhn:2001hz}
J.H. Kühn et~al.,
\newblock Nucl. Phys. B616 (2001) 286,
  \href{http://www.arXiv.org/abs/hep-ph/0106298}{{\tt hep-ph/0106298}},
\newblock 

\bibitem{Kuhn:2004em}
J.H. Kühn et~al.,
\newblock Phys. Lett. B609 (2005) 277,
  \href{http://www.arXiv.org/abs/hep-ph/0408308}{{\tt hep-ph/0408308}},
\newblock 

\bibitem{Kuhn:2005az}
J.H. Kühn et~al.,
\newblock Nucl. Phys. B727 (2005) 368,
  \href{http://www.arXiv.org/abs/hep-ph/0507178}{{\tt hep-ph/0507178}},
\newblock 

\bibitem{Kuhn:2005gv}
J.H. Kühn et~al.,
\newblock JHEP 03 (2006) 059,
  \href{http://www.arXiv.org/abs/hep-ph/0508253}{{\tt hep-ph/0508253}},
\newblock 

\bibitem{Beenakker:1993yr}
W. Beenakker et~al.,
\newblock Nucl. Phys. B411 (1994) 343,
\newblock 

\bibitem{Kao:1999kj}
C. Kao and D. Wackeroth,
\newblock Phys. Rev. D61 (2000) 055009,
  \href{http://www.arXiv.org/abs/hep-ph/9902202}{{\tt hep-ph/9902202}},
\newblock 

\bibitem{Kuhn:2005it}
J.H. Kühn, A. Scharf and P. Uwer,
\newblock Eur. Phys. J. C45 (2006) 139,
  \href{http://www.arXiv.org/abs/hep-ph/0508092}{{\tt hep-ph/0508092}},
\newblock 

\bibitem{Bernreuther:2005is}
W. Bernreuther, M. Fücker and Z.G. Si,
\newblock Phys. Lett. B633 (2006) 54,
  \href{http://www.arXiv.org/abs/hep-ph/0508091}{{\tt hep-ph/0508091}},
\newblock 

\bibitem{Moretti:2006nf}
S. Moretti, M.R. Nolten and D.A. Ross,
\newblock (2006), \href{http://www.arXiv.org/abs/hep-ph/0603083}{{\tt
  hep-ph/0603083}},
\newblock 

\bibitem{PaVe79}
G. Passarino and M. Veltman,
\newblock \npb B160 (1979) 151.

\bibitem{Denner:1991kt}
A. Denner,
\newblock Fortschr. Phys. 41 (1993) 307,
\newblock 

\bibitem{Jezabek:1993eh}
M. Jezabek and J.H. Kühn,
\newblock Prepared for Workshop on e+ e- Collisions, Hamburg, Germany, 2-3 Apr
  1993.

\bibitem{Bernreuther:2006xy}
W. Bernreuther, M. Fücker and Z.G. Si,
\newblock \href{http://www.arXiv.org/abs/hep-ph/0610xyz}{{\tt hep-ph/0610334}},
\newblock 

\bibitem{Kretzer:2003it}
S. Kretzer et~al.,
\newblock Phys. Rev. D69 (2004) 114005,
  \href{http://www.arXiv.org/abs/hep-ph/0307022}{{\tt hep-ph/0307022}},
\newblock 

\bibitem{Bonciani:1998vc}
R. Bonciani et~al.,
\newblock Nucl. Phys. B529 (1998) 424,
  \href{http://www.arXiv.org/abs/hep-ph/9801375}{{\tt hep-ph/9801375}},
\newblock 

\end{thebibliography}
\newcommand{\zp}{Z. Phys. }\def\as{\alpha_s }\newcommand{\prd}{Phys. Rev.
  }\newcommand{\pr}{Phys. Rev. }\newcommand{\prl}{Phys. Rev. Lett.
  }\newcommand{\npb}{Nucl. Phys. }\newcommand{\psnp}{Nucl. Phys. B (Proc.
  Suppl.) }\newcommand{\pl}{Phys. Lett. }\newcommand{\ap}{Ann. Phys.
  }\newcommand{\cmp}{Commun. Math. Phys. }\newcommand{\prep}{Phys. Rep.
  }\newcommand{\jmp}{J. Math. Phys. }\newcommand{\rmp}{Rev. Mod. Phys. }

\end{document}